\documentclass[aps,prd,twocolumn,superscriptaddress,nofootinbib,showkeys]{revtex4-2} 

\usepackage{amsmath}
\usepackage{amssymb}
\usepackage{physics}
\usepackage{siunitx}

\usepackage{graphicx}
\usepackage{lineno} 
\usepackage{booktabs}
\usepackage{hyperref}

\begin{document}

\title{Galactic Diffuse Gamma-Ray and Neutrino Emission from Cosmic-Ray Interactions in Stellar Atmospheres}

\author{Yanbo Wang}
\email{ybwang@pmo.ac.cn} 
\affiliation{Key Laboratory of Dark Matter and Space Astronomy \& Key Laboratory of Radio Astronomy, Purple Mountain Observatory, Chinese Academy of Sciences, Nanjing 210023, Jiangsu, China}
\affiliation{School of Astronomy and Space Science, University of Science and Technology of China, Hefei 230026, Anhui, China}

\author{Zhenglong Wang}
\affiliation{Key Laboratory of Dark Matter and Space Astronomy \& Key Laboratory of Radio Astronomy, Purple Mountain Observatory, Chinese Academy of Sciences, Nanjing 210023, Jiangsu, China}
\affiliation{School of Astronomy and Space Science, University of Science and Technology of China, Hefei 230026, Anhui, China}

\author{Rui Zhang}
\affiliation{Key Laboratory of Dark Matter and Space Astronomy \& Key Laboratory of Radio Astronomy, Purple Mountain Observatory, Chinese Academy of Sciences, Nanjing 210023, Jiangsu, China}

\author{Yi Zhang}
\email{Corresponding author: zhangyi@pmo.ac.cn}
\affiliation{Key Laboratory of Dark Matter and Space Astronomy \& Key Laboratory of Radio Astronomy, Purple Mountain Observatory, Chinese Academy of Sciences, Nanjing 210023, Jiangsu, China}
\affiliation{School of Astronomy and Space Science, University of Science and Technology of China, Hefei 230026, Anhui, China}
\date{\today}

\begin{abstract}
The Galactic diffuse gamma-ray emission is conventionally modeled as the product of cosmic-ray interactions with the interstellar medium. However, the cumulative contribution of stellar atmospheres acting as hadronic interaction targets remains an unexplored multi-messenger background. In this work, we present the first systematic evaluation of this stellar diffuse emission by coupling MESA stellar evolution profiles and magnetic-field-modulated cosmic-ray transport with a 3D Galactic population synthesis framework. We find that the cumulative stellar contribution to the Galactic diffuse gamma-ray flux is negligible at 1 TeV, and the associated diffuse neutrino flux ($\sim 10^{-17}\;\mathrm{TeV\;cm^{-2}\;s^{-1}\;sr^{-1}}$) remains orders of magnitude below current IceCube limits. Nevertheless, at ultra-high energies ($>10\;\mathrm{TeV}$), this emission establishes an irreducible local background that overtakes the strongly attenuated extragalactic isotropic gamma-ray background. Our results demonstrate that the Galactic stellar ensemble is a strictly sub-dominant background, indicating that stellar subtraction templates are not required for identifying Galactic PeVatrons or constraining dark matter annihilation.

\end{abstract}


\maketitle
\section{Introduction}
\label{sec:introduction}
The Galactic diffuse gamma-ray emission (GDGE) at TeV–PeV energies is conventionally attributed to cosmic-ray (CR) interactions with the interstellar medium (ISM)~\cite{Ackermann2012}. However, recent LHAASO measurements in the 1–100 TeV band reveal diffuse fluxes in the Galactic plane that exceed standard gas-model predictions by factors of 1.5 to 2.7, alongside significant morphological deviations from ISM templates~\cite{LHAASO2023}. These unresolved excesses strongly motivate the investigation of alternative production mechanisms and unexplored source populations within the Galaxy.

While historically treated merely as local CR absorbers, stars are now recognized as active high-energy emitters. Fermi-LAT has long detected steady GeV gamma-ray emission from the quiescent Sun~\cite{Fermi2011}, a phenomenon recently observed extending into the TeV regime by HAWC~\cite{HAWC2023}. Theoretically, sub-photospheric magnetic fields trap incident CR protons via magnetic mirroring~\cite{Ng2025}. This significantly enhances the effective target column density, effectively turning the stellar atmosphere into an optically thick target that produces very-high-energy gamma rays and all-flavor neutrinos via hadronic cascades ($pp \to \pi^0 \to 2\gamma$ and $pp \to \pi^\pm \to \nu$).

Extending this localized solar mechanism to the $\sim 10^{11}$ stars in the Milky Way requires evaluating their collective CR interaction capacity, which is governed jointly by stellar mass and magnetic confinement. Excluding the extended hot gas in the Galactic halo, stars dominate the Galactic baryonic mass budget, outweighing cold interstellar gas by a factor of 3 to 7~\cite{BlandHawthorn2016}. Although severe pair-production attenuation restricts escaping high-energy gamma rays to a shallow atmospheric skin, the efficiency of magnetic particle trapping fundamentally amplifies the CR interaction probability. The tight spatial correlation between stellar populations and the Galactic disk necessitates a quantitative assessment of whether the superposition of stellar atmospheres contributes meaningfully to the GDGE.

To date, the contribution of stellar atmospheric CR interactions to the high-energy diffuse background has remained virtually unexplored. While recent works have treated the Galactic stellar population as a macroscopic source of MeV thermal and nuclear neutrinos~\cite{Martinez2025}, we present the first systematic evaluation of their external hadronic interactions in the TeV–PeV band. In this work, we develop a multi-scale computational framework that couples internal stellar structural profiles—derived from Modules for Experiments in Stellar Astrophysics (MESA) simulations~\cite{MESA}—with magnetic-field-modulated CR transport models~\cite{Ng2025}. By integrating these microscopic yields into a 3D Galactic population synthesis model with fine mass binning and metallicity dependence, we systematically evaluate the cumulative multi-messenger (gamma-ray and neutrino) diffuse background from stars.

The remainder of this paper is organized as follows: Section II details the multi-scale computational framework. Section III presents the predicted radiation yields, spatial morphologies, and comparisons against current observational limits. Finally, Section IV discusses the physical implications for standard background and summarizes our conclusions.

\section{Methodology}
\label{sec:methodology}

To quantify the stellar atmospheric contribution to the GDGE and neutrino flux, we develop a multi-scale computational framework that bridges microscopic particle transport with Galactic-scale population synthesis. The methodology is structured into three primary stages:

First, we calculate the multi-messenger yields at the individual stellar level (Secs.~\ref{sec:cr_injection}--\ref{sec:stellar_sample}). By coupling stellar internal structural profiles from MESA simulations with hadronic interaction models, we evaluate the particle production efficiency. To account for the substantial uncertainties in stellar magnetospheres, we bracket the results using two limiting transport regimes: a field-free ballistic limit and a magnetic-field-modulated mirroring scenario.

Second, we implement a Galactic population synthesis model to transition from single-source emissivities to a macroscopic distribution (Sec.~\ref{sec:pop_synthesis}). This stage incorporates the Initial Mass Function (IMF), the age-metallicity relation (AMR), and the chemo-dynamical spatial structures of the Milky Way, including the bulge, thin disk, and thick disk.

Finally, we construct a continuous 3D volume emissivity grid through the superposition of the discrete stellar population. Integrating this emissivity along the line of sight (LoS) yields the predicted all-sky diffuse radiation maps.

\subsection{Cosmic-Ray Injection and Interaction Model}
\label{sec:cr_injection}
The production of secondary gamma rays and neutrinos is initiated by inelastic collisions between relativistic CR protons and stellar atmospheric nuclei. To accurately represent the incident flux $\Phi_p(E_p)$, we adopt a composite Local Interstellar CR Spectrum based on direct measurements from DAMPE~\cite{DAMPE2019} and high-altitude extensive air shower observations from LHAASO~\cite{LHAASO:2025proton}. This spectrum accounts for the observed spectral softening at $\sim 10$ TeV and the subsequent hardening at $\sim 0.34$ PeV before the proton knee at $\sim 3.3$ PeV. The spectrum is parameterized via a multi-break power law:

\begin{equation}
\begin{aligned}
\Phi_p(E_p) &= \Phi_0 \left( \frac{E_p}{1\,\mathrm{TeV}} \right)^{-\alpha_1} \times \\
&\quad \begin{cases}
1, & E_p < E_s \\
(E_p/E_s)^{\alpha_1 - \alpha_2}, & E_s \le E_p < E_h \\
\mathcal{C}_1 (E_p/E_h)^{\alpha_1 - \alpha_3}, & E_h \le E_p < E_k \\
\mathcal{C}_1 \mathcal{C}_2 (E_p/E_k)^{\alpha_1 - \alpha_4}, & E_p \ge E_k ,
\end{cases}
\end{aligned}
\end{equation}
where the continuity constants are defined as $\mathcal{C}_1 = (E_h/E_s)^{\alpha_1 - \alpha_2}$ and $\mathcal{C}_2 = (E_k/E_h)^{\alpha_1 - \alpha_3}$. The spectral indices are $\alpha_1 = 2.60$ and $\alpha_2 = 2.85$ for the low-energy regime~\cite{DAMPE2019}, transitioning to $\alpha_3 = 2.51$ and $\alpha_4 = 3.50$ at PeV energies~\cite{LHAASO:2025proton}. The corresponding transition energies are $E_s = 10\,\mathrm{TeV}$, $E_h = 340\,\mathrm{TeV}$, and $E_k = 3300\,\mathrm{TeV}$.

In the TeV–PeV regime, stellar high-energy emission is dominated by hadronic interactions. Leptonic contributions via inverse Compton (IC) scattering of cosmic-ray electrons on stellar radiation fields are severely suppressed due to the inherently lower primary electron flux and the transition into the deep Klein-Nishina regime~\cite{Zhou2017}. Consequently, the secondary emission spectrum closely traces the primary cosmic-ray proton distribution. Gamma rays are primarily generated through neutral pion decay:
\begin{equation}
pp \rightarrow \pi^0 + X \rightarrow 2\gamma + X
,\end{equation}
while the associated all-flavor neutrinos are produced via charged pion decay chains:
\begin{equation}
\begin{split}
pp \rightarrow \pi^+ + X &\rightarrow \mu^+ + \nu_\mu \rightarrow e^+ + \nu_e + \bar{\nu}_\mu + X \\
pp \rightarrow \pi^- + X &\rightarrow \mu^- + \bar{\nu}_\mu \rightarrow e^- + \bar{\nu}_e + \nu_\mu + X.
\end{split}
\end{equation}

To rigorously quantify the multi-messenger yields, we parameterize the total $pp$ inelastic cross section, $\sigma_{pp}(E_p)$, following Kafexhiu et al.~\cite{Kafexhiu2014}. The differential production cross sections for secondary gamma rays and neutrinos are computed using the AAfrag hadronic interaction package~\cite{AAfrag2019}. Convolving these differential cross sections with the effective proton column density accumulated in the stellar atmosphere provides a self-consistent calculation of the multi-messenger emission spectra from GeV to PeV energies.

\subsection{Field-Free Ballistic Limit}
\label{sec:ballistic}
In the limit of negligible magnetic fields, cosmic rays propagate along rectilinear trajectories through the stellar atmosphere. This ballistic approximation serves as a conservative lower bound for the multi-messenger yield.

\begin{figure}[t]
\centering
\includegraphics[width=0.48\textwidth]{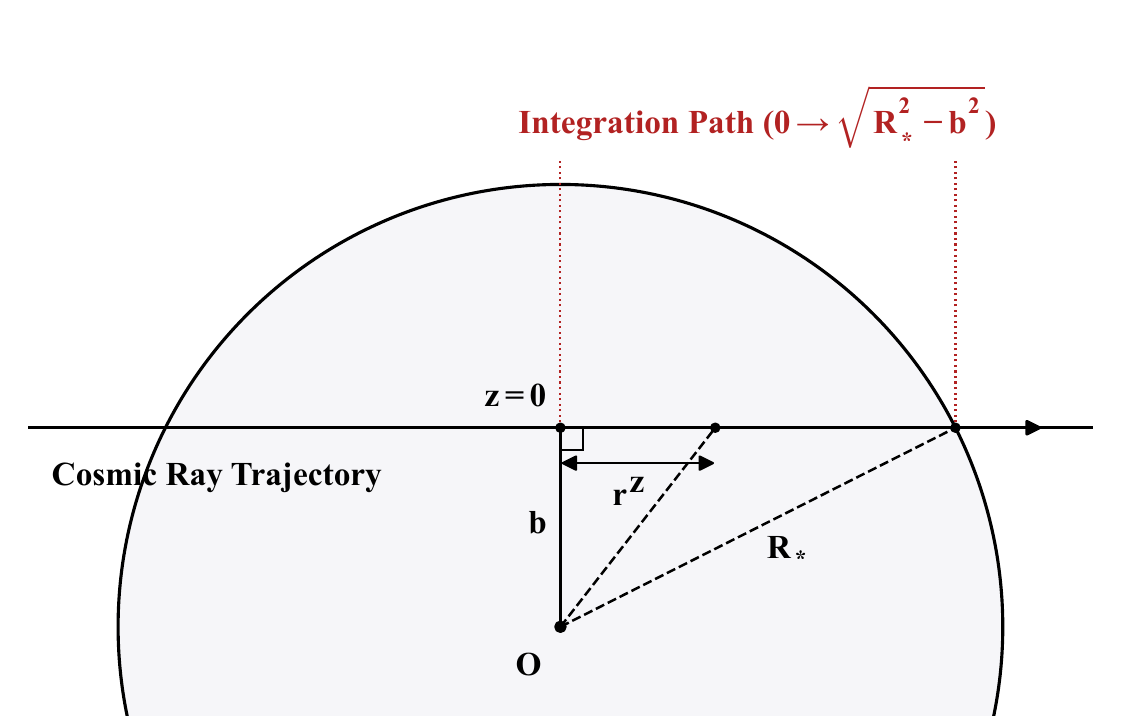}
\caption{Geometric configuration for cosmic-ray transport through a stellar atmosphere. A primary proton traverses along a ballistic trajectory (solid line) characterized by the impact parameter $b$. The total column density $T(b)$ is integrated from the symmetry point ($z=0$) to the stellar surface at $R_*$ over the path length $\sqrt{R_*^2 - b^2}$. The coordinate $r$ denotes the instantaneous radial position during propagation.}
\label{fig:geometry}
\end{figure}

The geometric setup for the cosmic-ray trajectory through the stellar atmosphere is illustrated in Fig.~\ref{fig:geometry}. For a star of radius $R_{\star}$ with a radial density profile $\rho(r)$ extracted from MESA models~\cite{MESA}, we define the impact parameter $b$ as the perpendicular distance from the incident cosmic-ray trajectory to the stellar center ($0 \le b \le R_{\star}$). The total column density $T(b)$ along the trajectory coordinate $z$ is evaluated as:
\begin{equation}
T(b) = 2 \int_{0}^{\sqrt{R_{\star}^2 - b^2}} \rho\left(\sqrt{b^2 + z^2}\right) dz .
\end{equation}

In the ballistic approximation, cosmic-ray protons traverse the stellar atmosphere, producing secondary gamma rays primarily via inelastic $p\text{-}p$ collisions. The differential gamma-ray luminosity $L_{\gamma}(E_{\gamma})$ is obtained by integrating over the impact parameter $b$ and the primary proton intensity $I_{p}(E_{p})$:
\begin{equation}
\begin{split}
L_{\gamma}(E_{\gamma}) &= 2\pi^2 \int_{0}^{R_{\star}} b \, db \int_{E_{\rm th}}^{\infty} dE_{p} \, I_{p}(E_{p}) \\
&\quad \times \mathcal{Y}_{\rm casc}(E_{\gamma}, E_{p}, b) \, P_{\rm esc}(E_{\gamma}, b) \, .
\end{split}
\end{equation}
Here, the hadronic cascade yield $\mathcal{Y}_{casc}$ convolves the local target nucleon density with the differential production cross section, weighted by the survival probability of the primary proton along the straight ballistic path. Conceptually, this term defines the field-free baseline for the primary cosmic-ray interaction probability, which is mathematically mapped to the effective hadronic optical depth, $\tau_{pp}$, in the subsequent magnetic mirroring framework (Sec.~\ref{sec:magnetic_mirroring}). The photon escape probability is defined by the attenuation optical depth, $P_{esc}(E_{\gamma},b) = e^{-\tau_{\gamma}}$, which is heavily dominated by $e^+e^-$ pair production at high energies.

By explicitly separating the production yield (which corresponds to $\tau_{pp}$) from the escape attenuation (parameterized by $e^{-\tau_{\gamma}}$), this ballistic formulation establishes the fundamental multi-messenger transport operators. In the field-free limit, both opacities are strictly evaluated along the geometric chord length of the trajectory. As detailed in Sec.~\ref{sec:magnetic_mirroring}, the introduction of sub-photospheric magnetic fields preserves these underlying interaction operators but evaluates them over magnetically elongated proton trajectories, thereby decoupling the effective hadronic grammage from the macroscopic stellar radius without invoking additional non-standard interaction channels.

Unlike gamma rays, which are subject to severe pair-production attenuation ($P_{\rm esc} = e^{-\tau_\gamma}$) and thus geometrically restricted to a shallow atmospheric skin, neutrinos undergo negligible exit attenuation in the sub-PeV regime ($P_{\rm esc} \approx 1$). Consequently, under this field-free ballistic limit, the differential neutrino luminosity $L_{\nu}(E_{\nu})$ scales directly with the primary proton interaction rate integrated along the full geometric traversal path. This establishes the baseline thick-target emission upon which the magnetic confinement effects in Sec.~\ref{sec:magnetic_mirroring} are built.

\subsection{Semi-analytic Magnetic Mirroring Model}
\label{sec:magnetic_mirroring}
To quantitatively evaluate the multi-messenger yields from stars with robust magnetospheres, we adopt the semi-analytic treatment developed by Ng et al.~\cite{Ng2025}. This framework circumvents the stochastic nature of full-scale Monte Carlo simulations by mapping cosmic-ray trajectories onto an effective interaction opacity, constrained by the internal stellar structure.

\subsubsection{Magnetic field parameterization based on energy equipartition}
\label{sec:B_param}

Observations indicate the presence of strong horizontal internetwork fields beneath the photosphere of the quiet Sun. To extrapolate this mechanism to stars across different masses and evolutionary stages, we assume the magnetic energy density reaches equipartition with the turbulent kinetic energy density of the stellar convective envelope. 

The sub-photospheric magnetic field strength, $B(r)$, is governed jointly by the local matter density, $\rho(r)$, and the convective velocity, $v_{\text{conv}}$ (both of which are extracted from the MESA structural profiles detailed in Sec.~\ref{sec:stellar_sample}):
\begin{equation}
B(r) = \sqrt{4\pi f_B \rho(r) v_{\text{conv}}^2} ,
\end{equation}
where $f_B$ is the partition fraction of magnetic energy density to fluid kinetic energy density. 

The parameter $f_B(M)$ dictates the dynamo efficiency. In low-mass stars with deep convective envelopes, the dynamo operates more efficiently, pushing the system toward full equipartition ($f_B \to 1$). We adopt the following piecewise parameterization:
\begin{equation}
f_B(M) =
\begin{cases}
1.0, & M \le 0.30 \, M_{\odot} \\
1.0 - \frac{3}{7}\left(\frac{M}{M_{\odot}} - 0.30\right), & 0.30 \, M_{\odot} < M \le 1.0 \, M_{\odot} \\
0.7, & 1.0 \, M_{\odot} < M \le 1.41 \, M_{\odot} \\
0.0, & M > 1.41 \, M_{\odot} .
\end{cases}
\end{equation}
For stars exceeding $1.41 \, M_{\odot}$, the convective envelope vanishes (the Kraft break~\cite{Beyer2024}), terminating the dynamo mechanism and truncating the magnetic field enhancement to zero.

\subsubsection{Particle trajectories and multi-messenger interaction efficiency}
\label{sec:particle_traj}

Under this magnetic configuration, cosmic-ray proton transport is governed by the Lorentz force, $d\vec{p}/dt = q(\vec{\beta}\times\vec{B})$, evaluated in the Gaussian unit system. Rather than executing full-cascade simulations, we employ analytic approximations to compute multi-messenger yields. For a proton moving along a specific trajectory, its accumulated hadronic interaction depth is $\tau_{pp} = \int \sigma_{pp} n(l) dl$.

\begin{figure}[htbp]
    \centering
    \includegraphics[width=\columnwidth]{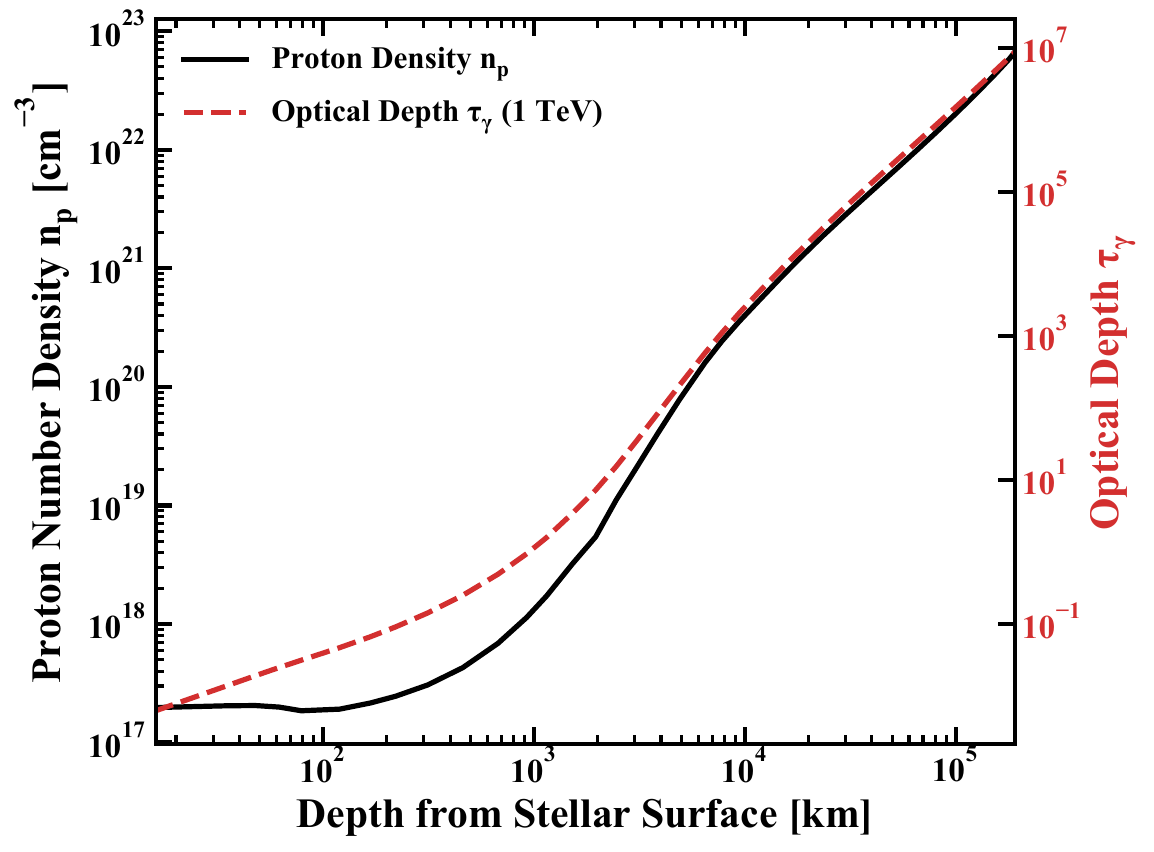}
    \caption{Radial profiles of the proton number density $n_p$ (solid black) and the corresponding 1 TeV $\gamma$-ray optical depth $\tau_\gamma$ (dashed red) as functions of depth from the stellar surface. The monotonic increase in target density drives the transition into the optically thick regime ($\tau_\gamma > 1$) within the sub-photospheric layers.}
    \label{fig:solar_profile}
\end{figure}

Following the transport dichotomy established in Sec.~\ref{sec:ballistic}, the observable gamma-ray production efficiency, $\eta_\gamma$, is heavily suppressed by pair-production absorption ($\gamma + Z \to e^+ + e^- + Z$) along the exit trajectory:
\begin{equation}
\eta_\gamma(E_{p},\Omega) \approx \tau_{pp} \cdot e^{-\tau_{\gamma}} ,
\end{equation}
where $\tau_{\gamma}$ is the electromagnetic optical depth along the exit path. Conversely, the stellar envelope's transparency to sub-PeV neutrinos allows their production and escape efficiency to scale directly with the primary proton interaction probability along the magnetically trapped trajectory:
\begin{equation}
\eta_\nu(E_{p},\Omega) \approx 1 - e^{-\tau_{pp}} .
\end{equation}
Thus, in the optically thick regime ($\tau_{pp} \gg 1$), the neutrino luminosity inherently exceeds the surface-constrained gamma-ray luminosity, preserving the thick-target divergence without requiring full cascade simulations.

\subsubsection{Total flux integration}
\label{sec:flux_integration}

The differential gamma-ray flux from a single star, $d\Phi_\gamma/dE_\gamma$, is derived by convolving the incident cosmic-ray proton intensity, $I_p(E_p)$, with the geometric surface filling factor of the magnetic field, $F_{\text{sur}}$:
\begin{equation}
\frac{d\Phi_\gamma}{dE_\gamma} = F_{\text{sur}} \cdot N_{\text{nuc}} \cdot \int dE_p I_p(E_p) \int d\Omega \frac{\eta_\gamma(E_p, \Omega)}{\kappa_\pi} ,
\end{equation}
where $N_{\text{nuc}} \approx 1.8$ is the nuclear enhancement factor accounting for heavier cosmic-ray species, and $\kappa_\pi$ represents the fractional energy transferred to the pion channel (inelasticity). To account for internal structure and age-driven magnetic braking, $F_{\mathrm{sur}}$ is parameterized as the product of a structural factor, $f_{\mathrm{struct}}(M)$, and a temporal decay factor, $f_{\mathrm{age}}(t, M)$:
\begin{equation}
F_{\mathrm{sur}}(M, t) = \min\Big[ f_{\mathrm{struct}}(M) \cdot f_{\mathrm{age}}(t, M), \, 1.0 \Big] .
\end{equation}

The structural factor $f_{\mathrm{struct}}(M)$ maps the depth of the convective envelope. The mass nodes ($0.35 \, M_{\odot}$ and $0.50 \, M_{\odot}$) trace the topological transitions observed in low-mass stars crossing the fully convective boundary~\cite{See2019}. We utilize a smooth step function, $S(x) = 0.5 + 0.5 \cos(\pi x)$ defined for $x \in [0, 1]$, to connect these physical regimes. Defining the dimensionless mass $m = M/M_{\odot}$:
\begin{equation}
f_{\mathrm{struct}}(m) =
\begin{cases}
0.95, & m \le 0.35 \\[1ex]
0.45 + 0.5 S\!\left(\frac{m-0.35}{0.15}\right), & 0.35 < m \le 0.5 \\[1ex]
0.3 + 0.15 S\!\left(\frac{m-0.5}{0.35}\right), & 0.5 < m \le 0.85 \\[1ex]
0.3, & 0.85 < m \le 1.32 \\[1ex]
0.3 S\!\left(\frac{m-1.32}{0.09}\right), & 1.32 < m \le 1.41 \\[1ex]
0, & m > 1.41 .
\end{cases}
\end{equation}
This bounds the saturated surface coverage of fully convective stars at the low-mass end~\cite{See2019} and truncates at the Kraft break at the high-mass end~\cite{Beyer2024}.

The age decay factor, $f_{\mathrm{age}}(t, M)$, obeys the Skumanich relation. Following a mass-dependent saturation age, $t_{\mathrm{sat}}(M)$~\cite{Johnstone2021}, magnetic activity decays as a power law:
\begin{equation}
f_{\mathrm{age}}(t, M) =
\begin{cases}
1.0, & t \le t_{\mathrm{sat}}(M) \\
\left( \frac{t}{t_{\mathrm{sat}}(M)} \right)^{-0.5}, & t > t_{\mathrm{sat}}(M) .
\end{cases}
\end{equation}

\begin{figure}[htbp]
    \centering
    \includegraphics[width=1\columnwidth]{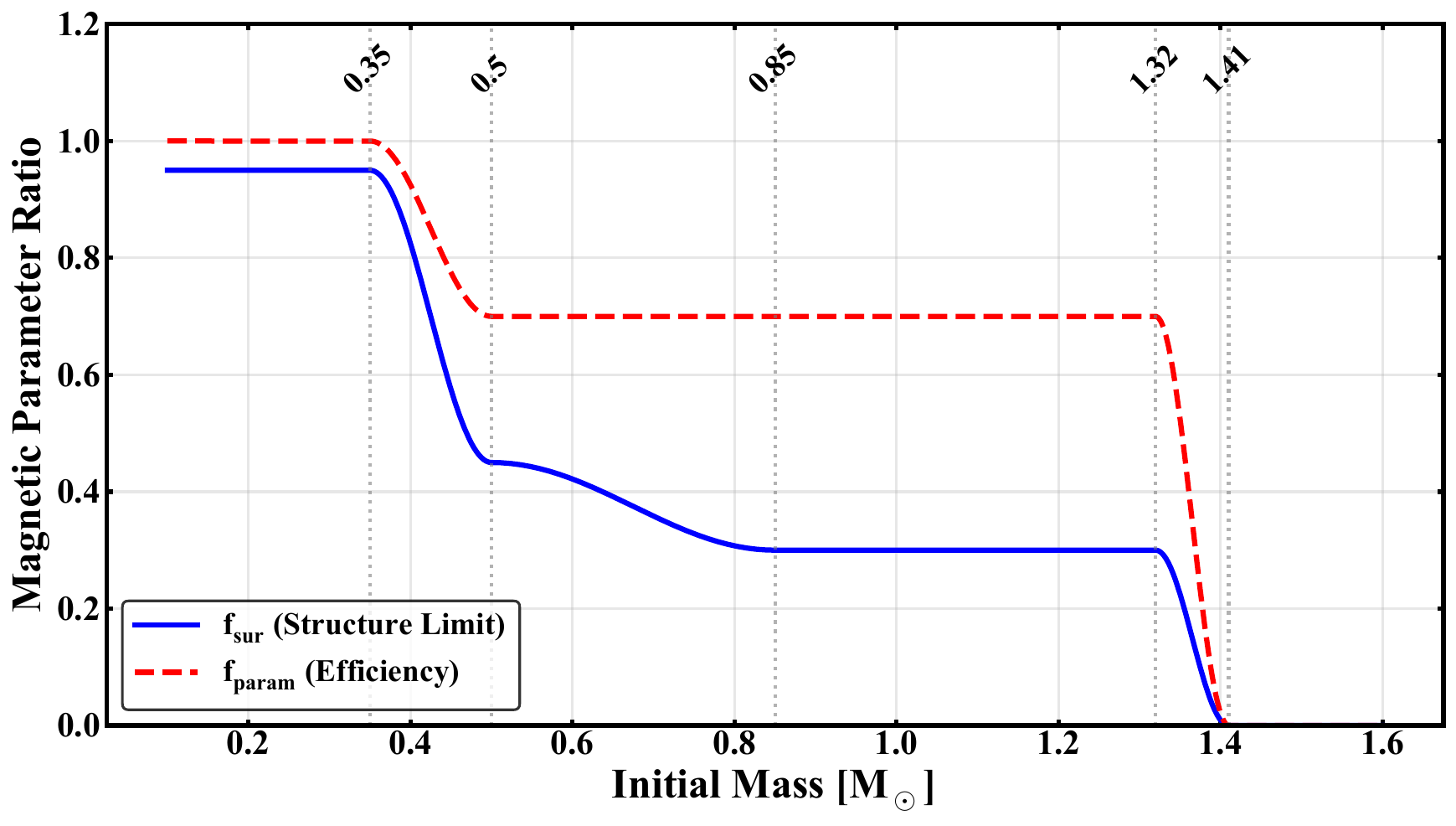}
    \caption{
    Mass-dependence of the local magnetic field parameters for main-sequence stars. The solid blue line represents the surface magnetic filling factor $f_{\text{struct}}$, while the dashed red line shows the energy partition fraction $f_B$. The parameterization accounts for the saturated dynamo efficiency in low-mass, fully convective stars ($M \leq 0.35 M_\odot$) and the abrupt truncation at the Kraft break ($M \approx 1.41 M_\odot$).~\cite{Beyer2024}.}
    \label{fig:magnetic_parameters}
\end{figure}

The dependence of these structural parameters on initial mass is illustrated in Fig.~\ref{fig:magnetic_parameters}. In this framework, magnetic mirroring operates exclusively during the main-sequence phase. For post-main-sequence evolution (e.g., the red giant branch), severe envelope expansion destroys the dynamo mechanism. The code rigorously enforces $F_{\mathrm{sur}} = 0$ and $f_B = 0$ for these stages, automatically collapsing the transport back to the field-free ballistic limit.

\subsection{Construction of the Stellar Profile Library}
\label{sec:stellar_sample}

To model Galactic-scale stellar multi-messenger emission, we compute interaction yields directly from internal stellar structural profiles rather than relying on macroscopic scaling relations. 
To construct a comprehensive stellar database, we employ a composite approach. For low-mass populations ($M \le 1.0\,M_{\odot}$), we perform dedicated simulations using the MESA code~\cite{MESA} to generate fine-grained internal profiles. For intermediate and massive stars ($M > 1.0\,M_{\odot}$), we utilize the highly validated MESA evolution models provided by the NuGrid dataset~\cite{Ritter2018}. 
MESA theoretical isochrones for solar-like scales have been robustly validated against high-precision observations (e.g., Gaia EDR3 Hyades data)~\cite{Brandner2023}, ensuring the physical reliability of the atmospheric inputs in our model.

\subsubsection{Evolutionary Grid and Parameter Space}
\label{sec:evo_grid}

Our baseline dataset encompasses the primary components of the Galactic stellar population. The benchmark structural library spans initial masses ($M_{\mathrm{ini}}$) from $0.1\,M_{\odot}$ to $25\,M_{\odot}$ on a representative solar-metallicity grid ($Z=0.02$). The output captures the complete evolutionary history from the zero-age main sequence (ZAMS) to the evolutionary endpoint (white dwarf cooling or pre-supernova).

The mass parameter space is strictly truncated at $25\,M_{\odot}$. This cutoff is based on rigorous physical considerations, consistent with recent Galactic population synthesis frameworks~\cite{Martinez2025}. Although massive stars ($M > 25\,M_{\odot}$) possess large geometric cross sections ($\sigma_{\rm geo} \propto R^2 \propto M^{1.6}$), their contribution to the total Galactic diffuse flux is doubly suppressed by the IMF and their brief evolutionary lifetimes. 

Under the assumption of a steady-state Galactic star formation rate, the present-day number density of stars, $n(M)$, is proportional to $\xi_{\mathrm{IMF}}(M) \times \tau_{\mathrm{life}}(M)$. Given the power-law decline of the IMF at the high-mass end ($\propto M^{-2.3}$) and the rapid shortening of main-sequence lifetimes ($\tau \propto M^{-2.5}$), the population density steeply drops as $n(M) \sim M^{-4.8}$. This severe statistical depletion far outpaces the growth of their geometric cross sections. Quantitative estimates demonstrate that the flux contribution per unit mass interval asymptotically saturates near $3\,M_{\odot}$. Therefore, the cumulative contribution from stars above $25\,M_{\odot}$ is statistically negligible. Omitting this regime rigorously bounds the systematic uncertainties introduced by highly non-linear extreme-mass evolutionary models without compromising the macroscopic diffuse yield.

\subsubsection{Stage-based Temporal Sampling Strategy}
\label{sec:sampling_strategy}

Stellar evolution spans timescales differing by several orders of magnitude—ranging from billions of years on the main sequence to mere thousands during thermal pulses on the asymptotic giant branch (AGB). Consequently, uniform sampling across numerical time steps yields severe data redundancy during quiescent phases and drastically undersamples rapid structural transitions. To mitigate this, we develop an automated stage-identification and time-weighted sampling algorithm that couples internal structural profiles with surface time-series data. 

Following standard stellar evolution theory~\cite{Kippenhahn2012}, we partition the continuous evolutionary tracks into five discrete physical stages. The boundary criteria for this classification are summarized in Table~\ref{tab:evo_stages}.

\begin{table*}[t] 
    \caption{\label{tab:evo_stages} Physical criteria for stellar evolutionary stage identification.}
    \begin{ruledtabular} 
        \begin{tabular}{lll}
            \textrm{\textbf{Evolutionary Stage}} & \textrm{\textbf{Dominant Process}} & \textrm{\textbf{Physical Criterion / Boundary}} \\
            \colrule 
            Main Sequence (MS) & Core H burning & Central H fraction $X_{\mathrm{H}} > 10^{-4}$~\cite{Ekstrom2012} \\ 
            Hertzsprung Gap \& Red Giant Branch (RGB) & Shell H burning & Central He fraction $X_{\mathrm{He}} > 0.90$, $\log T_c < 8.0$~\cite{Kippenhahn2012} \\ 
            Core He Burning (CHeB) & Core He burning & $X_{\mathrm{He}} > 10^{-4}$ and $\log T_c \ge 8.0$ \\ 
            Early AGB (E-AGB) & Quiescent C-O core & Central mass fractions $X_{\mathrm{C}} + X_{\mathrm{O}} \ge 0.90$~\cite{Iben1983} \\ 
            Thermally Pulsing AGB (TP-AGB) & He shell flashes (PDCZ) & Diffusion coeff. $D_{\mathrm{coeff}} > 10^{12}\ \mathrm{cm}^2\ \mathrm{s}^{-1}$~\cite{Iben1983, Herwig2005, Karakas2007} \\ 
        \end{tabular}
    \end{ruledtabular}
\end{table*}

To suppress numerical artifacts from MESA's adaptive stepping, we uniformly resample each stage into a high-density ensemble of snapshots at equal physical time intervals, ensuring the sampling cadence is well below the relevant stellar dynamical timescales and providing an unbiased basis for the subsequent population synthesis.

This adaptive sampling strategy ensures that brief but dynamically violent phases, such as the TP-AGB—characterized by intense convection and vigorous \textit{s}-process nucleosynthesis~\cite{Gallino1998, Pignatari2016}—receive appropriate statistical representation in the final structural grid. Such consistency is strictly required to compute unbiased macroscopic multi-messenger yields.

\subsubsection{Extraction of Internal Structural Profiles}
\label{sec:profile_extraction}

For each temporal snapshot, we extract high-resolution radial structural profiles. While previous macroscopic models relied exclusively on photospheric boundary conditions, quantifying sub-photospheric magnetic field enhancement mandates deep interior dynamical parameters. Consequently, we extract two primary radial distribution functions per model: the mass density, $\rho(r)$, which defines the target material profile for cosmic-ray hadronic interactions, and the convective velocity, $v_{\mathrm{conv}}(r)$. Both profiles are directly extracted from the MESA simulation results \cite{MESA}, with the latter being computed according to the standard Mixing Length Theory (MLT). Assuming turbulent energy equipartition, $v_{\mathrm{conv}}(r)$ governs the upper limit of the local magnetic field strength ($B \propto \sqrt{\rho v_{\mathrm{conv}}^2}$).

The finalized structural library encompasses approximately $10^5$ independent stellar models. Each is parameterized as a standardized lookup table of $r$, $\rho(r)$, and $v_{\mathrm{conv}}(r)$. These discrete radial profiles serve as the direct physical inputs for the subsequent semi-analytic particle transport calculations.

\subsection{Galactic Population Synthesis and Volume Emissivity Grid}
\label{sec:pop_synthesis}

We employ a grid-based population synthesis framework to extrapolate microscopic single-star interaction yields to macroscopic Galactic scales. The multi-messenger luminosity of a stellar atmosphere is governed by the stellar mass, evolutionary stage, and local cosmic-ray environment; thus, we construct a comprehensive three-dimensional spatial distribution of the stellar volume emissivity across the Milky Way.

\subsubsection{Galactic Spatial Structure and Stellar Density Distribution}
\label{sec:gal_structure}
To define a self-consistent three-dimensional spatial distribution for population sampling and absolute flux calibration, we partition the Milky Way into three primary structural components: the bulge, the thin disk, and the thick disk, adopting the parameterized stellar mass distributions from recent studies~\cite{Martinez2025}. The bulge density adopts a power-law profile with an exponential truncation:

\begin{equation}
    \rho_{b} = \frac{\rho_{0,b}}{(1 + r'/r_0)^{\alpha_b}} \exp\left[ -\left( \frac{r'}{r_{\mathrm{cut}}} \right)^2 \right] ,
\end{equation}
where the effective elliptical radius is $r' = \sqrt{R^2 + (z/q_b)^2}$. The structural parameters are $\rho_{0,b} = 101.0\,M_\odot\,\mathrm{pc}^{-3}$, $q_b = 0.5$, $\alpha_b = 1.8$, $r_0 = 75\,\mathrm{pc}$, and $r_{\mathrm{cut}} = 2.1\,\mathrm{kpc}$.
The disk components are governed by double-exponential profiles in radial and vertical coordinates. For the thin disk:
\begin{equation}
    \rho_{\mathrm{thin}} = \frac{\Sigma_{0,\mathrm{thin}}}{2 z_{d,\mathrm{thin}}} \exp\left( -\frac{|z|}{z_{d,\mathrm{thin}}} - \frac{R}{R_{d,\mathrm{thin}}} \right) ,
\end{equation}
where the local surface density is $\Sigma_{0,\mathrm{thin}} = 1070.0\,M_\odot\,\mathrm{pc}^{-2}$, the scale height is $z_{d,\mathrm{thin}} = 300.0\,\mathrm{pc}$, and the scale length is $R_{d,\mathrm{thin}} = 2.43\,\mathrm{kpc}$. The thick disk utilizes an identical functional form parameterized for a sparser, vertically extended older population: $\Sigma_{0,\mathrm{thick}} = 113.0\,M_\odot\,\mathrm{pc}^{-2}$, $z_{d,\mathrm{thick}} = 900.0\,\mathrm{pc}$, and $R_{d,\mathrm{thick}} = 3.88\,\mathrm{kpc}$.

To translate the macroscopic mass distribution into the actual spatial stellar number density, $n(R, z)$, we scale the total mass density by a globally averaged mean stellar mass, $M_{\mathrm{avg}} = 0.41\,M_\odot$~\cite{Kirkpatrick2024}:

\begin{equation}
    n(R, z) = \frac{\rho_b + \rho_{\mathrm{thin}} + \rho_{\mathrm{thick}}}{M_{\mathrm{avg}}} .
\end{equation}

\begin{figure*}[t]
    \centering
    \includegraphics[width=0.78\textwidth]{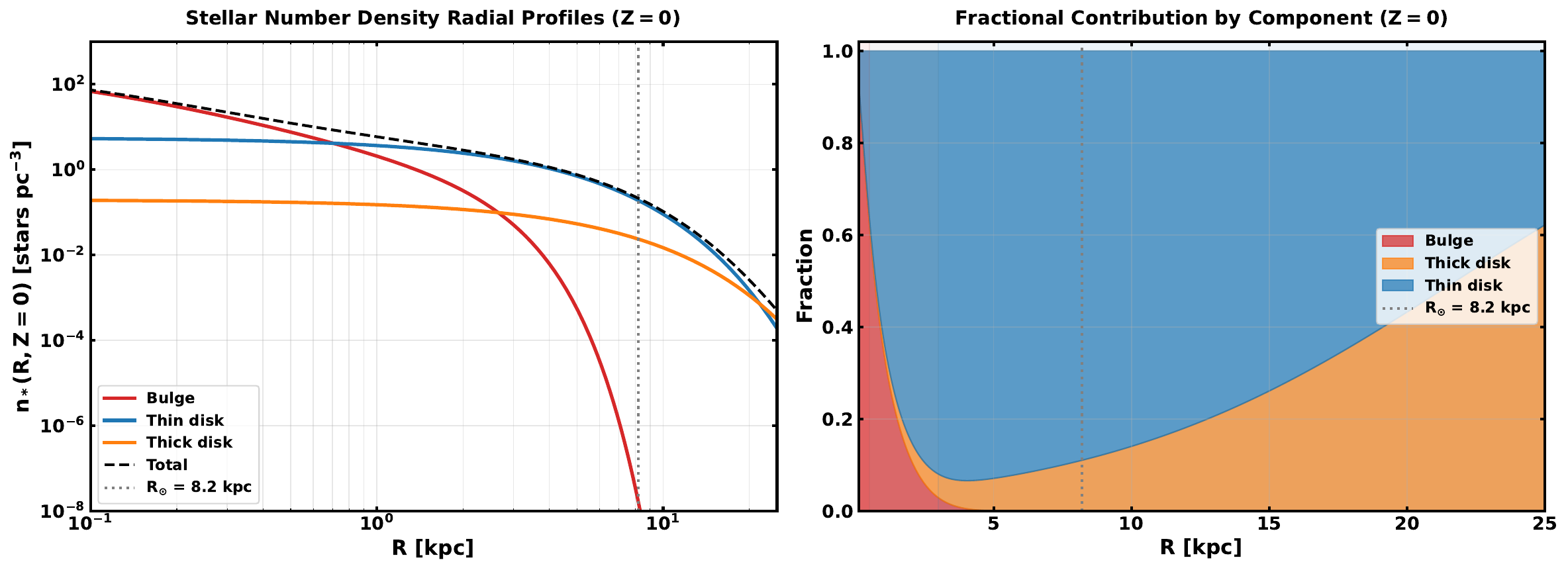}
    \caption{Radial profiles of the synthetic Galactic stellar population at the mid-plane ($Z=0$). Left: Stellar number density for the bulge, thin disk, and thick disk. Right: Fractional contribution of each component. The bulge dominates the inner kiloparsec, transitioning to a thin-disk-dominated regime at larger radii. Although the thick-disk fraction steadily increases toward the outer Galaxy, it remains physically subdominant within the integration grid.}
    
    \label{fig:radial_number_density}
\end{figure*}

Our baseline Galactic model superimposes a bulge, a thin disk, and a thick disk, omitting the low-density stellar halo. Line-of-sight integrations are bounded within a cylindrical grid ($R \le 15$ kpc, $|Z| \le 3$ kpc). As illustrated in Fig.~\ref{fig:radial_number_density} , the bulge dominates the inner kiloparsec, whereas the extended thin disk dictates the total Galactic stellar diffuse emission, including the solar neighborhood ($R_\odot = 8.2$ kpc).


\subsubsection{Random Sampling of Stellar Physical Properties}
\label{sec:stellar_sampling}
At each grid point, the simulated stars are distributed among the Galactic components (e.g., thin disk, thick disk, and bulge) in strict accordance with their respective local number densities. Physical properties are then assigned to trace the chemo-dynamical evolution of the Milky Way. Stellar masses are sampled from a broken power-law IMF:

\begin{equation}
\xi(M) \propto M^{-\alpha}, \quad \alpha = \begin{cases}
1.3, & 0.08 \, M_{\odot} \le M < 0.5 \, M_{\odot} \\
2.3, & 0.5 \, M_{\odot} \le M < 100 \, M_{\odot}
.\end{cases}
\end{equation}
While the IMF extends to $100 \, M_{\odot}$, our Monte Carlo population synthesis restricts the structural mass grid to $0.1\text{--}25 \, M_{\odot}$. The population synthesis effectively integrates the IMF down to the hydrogen-burning limit of $0.08 \, M_{\odot}$. For the extreme low-mass regime ($0.08\text{--}0.1 \, M_{\odot}$), we conservatively assign the multi-messenger yield of the $0.1 \, M_{\odot}$ fully convective model, as internal magnetic topologies are highly degenerate near this limit. 

The high-mass truncation at $25 \, M_{\odot}$ is strictly motivated by statistical and theoretical constraints. Macroscopically, stars with $M > 25 \, M_{\odot}$ constitute only $\sim 0.12\%$ of the total population; their contribution to the steady-state Galactic diffuse emission is severely suppressed by their exceedingly short main-sequence lifetimes. Microscopically, 1D spherically symmetric stellar models face severe physical limitations in this extreme-mass regime. Near the Eddington limit, wind mass-loss rates become highly unstable, and their core-collapse supernova (CCSN) phases lack self-consistent explosion mechanisms, often requiring artificial engines that fail to capture massive fallback effects. Truncating at $25 \, M_{\odot}$ safely encompasses the dominant effective target mass for Galactic diffuse emission while rigorously excluding the systematic biases and exponential computational costs introduced by unresolved non-linear processes.

Age and metallicity distributions follow distinct evolutionary trajectories for each structural component. The thin disk age follows an exponential distribution ($\Gamma_{\rm thin} = 0.12$). Its metallicity, $[{\rm Fe/H}]$, is governed by an AMR and spatial gradients:
\begin{equation}
    [{\rm Fe/H}] = \text{AMR}_{0} + k \, t_{\mathrm{age}} + \frac{\partial[{\rm Fe/H}]}{\partial R} (R - R_\odot) + \frac{\partial [{\rm Fe/H}]}{\partial |Z|} |Z|
,\end{equation}
where $k = -0.04$~dex/Gyr is the temporal evolution slope. Following Feuillet et al.~\cite{Feuillet2019}, the radial gradient is set to $\frac{\partial[{\rm Fe/H}]}{\partial R} = -0.059$~dex/kpc, and $\text{AMR}_{0} = 0.15$.

The thick disk comprises an older stellar population, modeled with a Gaussian age distribution ($11.0 \pm 1.5$~Gyr) and a fixed mean metallicity of $-0.5$~dex. 

The bulge is primarily composed of ancient stellar populations (mean age $11.0$~Gyr). Its metallicity adopts the bimodal distribution reviewed by Gonzalez and Gadotti~\cite{Gonzalez2015}: a metal-rich component ($[{\rm Fe/H}] \approx +0.3$) comprising 75\% of the mass, and a metal-poor component ($[{\rm Fe/H}] \approx -0.3$) comprising 25\%. Additionally, following the morphological constraints of Simion et al.~\cite{Simion2017}, we inject a young stellar population ($\sim 5\%$, age 2--5~Gyr) into the central bulge region ($|Z| \le 0.25$~kpc, $2.0 \le R \le 3.5$~kpc).

\subsubsection{Grid-based Average Spectral Construction}
\label{sec:grid_spectra}

Because the multi-messenger emission characteristics of an individual star are strictly governed by its mass ($M$), metallicity ($Z$), and evolutionary stage, we employ a $k$-nearest neighbor ($k$-NN) weighted interpolation algorithm to synthesize the local average emissivity at each spatial grid point.

For each sampled star in the population, an evolutionary survival criterion is first applied. Utilizing a lifetime interpolation grid parameterized by mass and metallicity, we evaluate the maximum stellar lifespan. Stars whose sampled ages exceed this limit are classified as stellar remnants (e.g., white dwarfs, neutron stars, or black holes) and are assigned zero multi-messenger yield.

To mitigate numerical fluctuations from discrete structural templates and correctly account for the physical duration of evolutionary phases, we introduce a temporal weighting mechanism, $\Delta t$. Surviving stars are mapped into a three-dimensional feature space ($\log M$, $\log Z$, $\log \text{Age}$). The $K=10$ nearest template neighbors are retrieved from the structural library via a cKDTree search. Distance weights are computed and convolved with the evolutionary duration factor, $\Delta t$, of each template to construct the normalized average differential luminosity per star, $\bar{L}(E)$:
\begin{equation}
    \bar{L}(E) = \frac{1}{N} \sum_{i=1}^{N_{\rm alive}} \sum_{j=1}^{K} w_{i,j} L_{j}(E)
,\end{equation}
where $N$ is the total number of sampled stars in the local grid, $N_{\rm alive}$ is the subset of living stars, $L_{j}(E)$ is the differential luminosity of the $j$-th discrete template, and $w_{i,j}$ represents the normalized combined weight of the inverse Euclidean distance and $\Delta t$. This mean differential luminosity serves as the core physical input for the subsequent macroscopic Galactic line-of-sight integration.

\subsubsection{Line-of-Sight Integration and Regional Spectral Synthesis}
\label{sec:los_integration}

To bridge the microscopic interaction yields with macroscopic observables, we evaluate the specific intensity, $I(E,l,b)$, which represents the differential flux per unit solid angle along a given directional coordinate $(l,b)$. We parameterize the line of sight (LoS) pointing in the direction of Galactic coordinates $(l, b)$ using the distance $s$ from the Sun. Along this path, the local Galactocentric cylindrical coordinates are mapped as $R_s \equiv R(s,l,b)$ and $z_s \equiv z(s,l,b)$.
In the optically thin limit, the geometric $1/s^2$ attenuation of individual stellar fluxes is analytically canceled by the $s^2$ expansion of the differential volume element $(dV=s^2\,d\Omega\,ds)$. The observable specific intensity follows from the linear integration of the local volume emissivity along the LoS:
\begin{equation}
I(E,l,b)=\frac{1}{4\pi}
\int_{0}^{s_{\rm grid}(l,b)}
n(R_s,z_s)\,
\bar{L}(E,R_s,z_s)\,ds .
\end{equation}
\noindent Here, $s_{\rm grid}(l,b)$ denotes the distance at which the LoS exits our synthesis grid ($0\le R\le 15~{\rm kpc}$, $|Z|\le 3~{\rm kpc}$).

To construct the total spectral energy distribution (SED) for a macroscopic Region of Interest (ROI), such as the Galactic plane or the high-latitude sky, we integrate the specific intensity over the corresponding solid angle. Accounting for the spherical Jacobian, the total differential energy flux, $E^2 dN/dE$, is defined as:
\begin{equation}
    E^2 \frac{dN}{dE} = E^2 \int_{\rm ROI} I(E, l, b) \cos b \, dl \, db .
\end{equation}
In our 3D population synthesis framework, this continuous integral is numerically evaluated across the discrete spatial grid by summing the specific intensities weighted by their true projected solid angles, $\Delta \Omega \approx \cos b \, \Delta l \Delta b$. This approach rigorously preserves the underlying spherical geometry and mitigates the latitude-dependent projection distortions inherent in uniform grid sampling. This unified macroscopic integration is applied simultaneously to both the gamma-ray and neutrino channels.


\section{Results}
\label{sec:results}
In this section, we quantitatively evaluate the contribution of Galactic stellar atmospheres to the diffuse gamma-ray and neutrino backgrounds, utilizing the physical model and simulation framework developed in Sec.~\ref{sec:methodology}. We begin by examining the microscopic multi-messenger yields of individual stars, scale these to macroscopic Galactic emission morphologies, and finally confront our predictions with existing VHE and UHE observations to constrain the stellar contribution.

\subsection{Stellar Physical Structure and Microscopic Interaction Processes}
\label{sec:results_micro}

\begin{figure}[htbp]
    \centering
    \includegraphics[width=0.95\columnwidth]{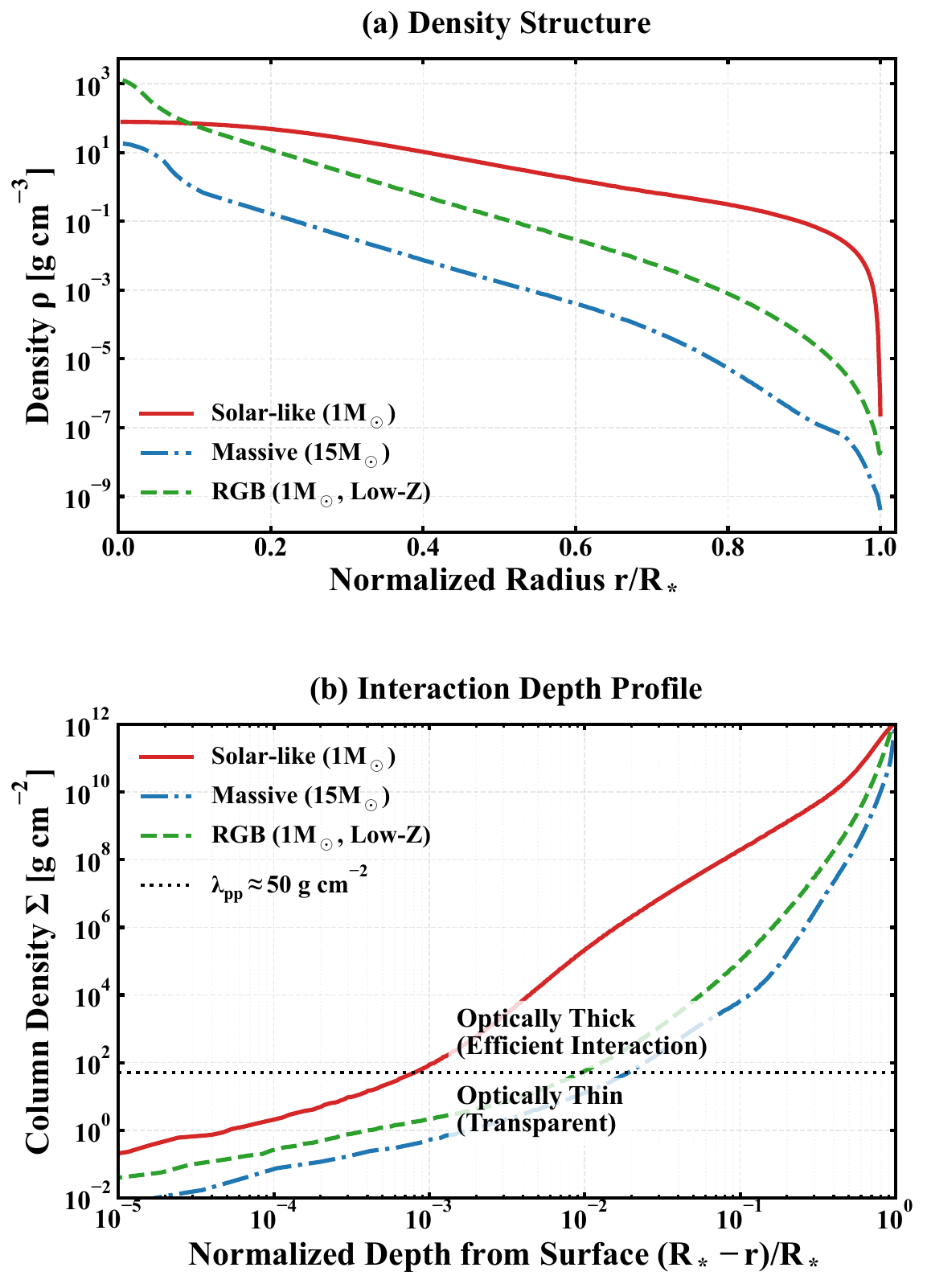}
    \caption{Internal structural profiles derived from MESA simulations for representative stellar types: solar-like ($1.0 M_\odot$, red), massive ($15 M_\odot$, blue), and red giant (RGB, green). (a) Radial mass density distribution $\rho(r)$. (b) Cumulative column density $\Sigma(r)$ integrated inward from the surface. The horizontal dashed line indicates the characteristic hadronic interaction length $\lambda_{pp} \approx 50\;\mathrm{g\;cm^{-2}}$, defining the threshold for efficient secondary production.}
    \label{fig:density_structure}
\end{figure}

\begin{figure}[htbp]
    \centering
    \includegraphics[width=0.95\columnwidth]{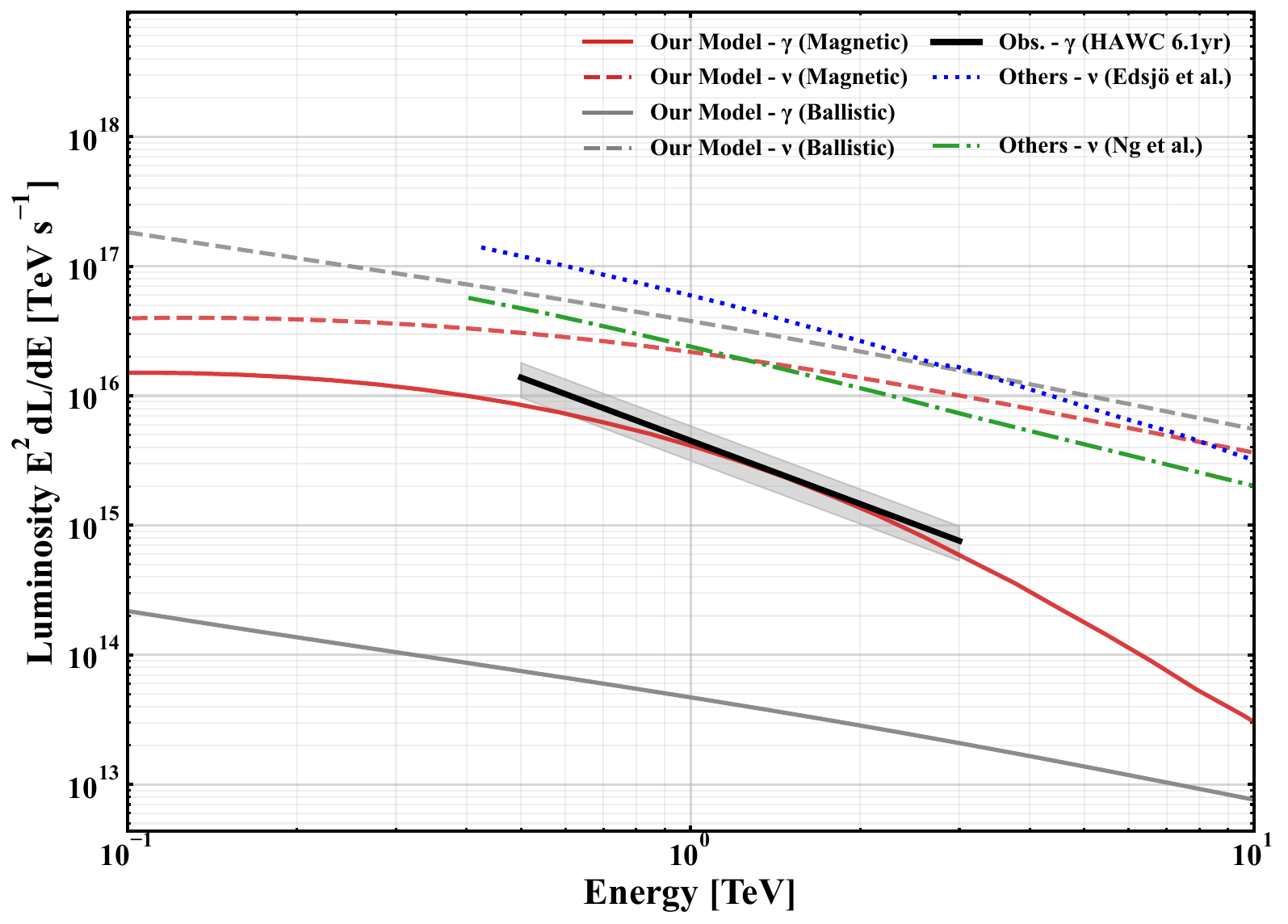}
    \caption{Spectral energy distribution (SED) of secondary gamma rays (solid) and all-flavor neutrinos (dashed) for the solar model. The magnetically modulated scenario (red) is contrasted with the field-free ballistic limit (gray). The shaded band represents the observationally-constrained TeV emission from the quiescent Sun by HAWC~\cite{HAWC2023}, providing the empirical calibration for the magnetic mirroring framework. For comparison, the predicted solar atmospheric neutrino fluxes from previous studies by Edsjö et al.~\cite{Edsjo2017} and Ng et al.~\cite{Ng2017} are also shown.}
    \label{fig:multimessenger_spectra}
\end{figure}

\begin{figure}[htbp]
    \centering
    \includegraphics[width=0.95\columnwidth]{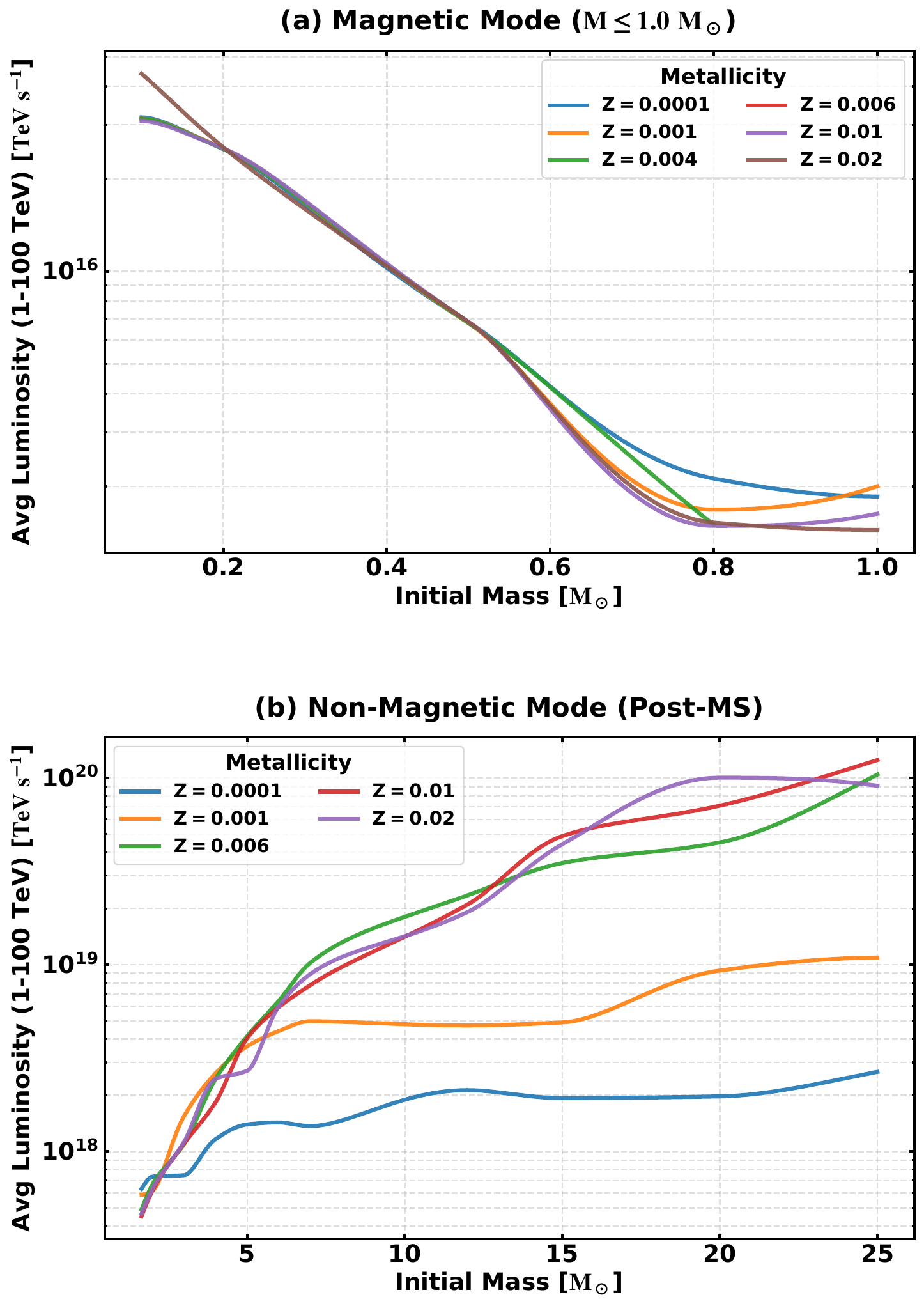}
    \caption{Time-weighted average gamma-ray luminosity (1–100 TeV) as a function of initial mass and metallicity $Z$. (a) Magnetic modulation regime ($M \leq 1.41 M_\odot$), showing the anti-correlation between luminosity and stellar mass due to convective dynamo efficiency. (b) Post-main-sequence ballistic regime, where emission scales with the geometric cross-section, exhibiting high sensitivity to $Z$-dependent envelope expansion.}
        
    \label{fig:lum_mass_relation}
\end{figure}

To evaluate the efficiency of stellar atmospheres as cosmic-ray targets, we extract internal structural profiles for representative stars across different evolutionary stages using MESA. Figure~\ref{fig:density_structure}(a) presents the radial density distributions, $\rho(r)$, for a solar-like star ($1.0\,M_\odot$), a massive main-sequence star ($15\,M_\odot$), and a low-metallicity red giant ($1\,M_\odot$). Because the internal density rises sharply with decreasing radius, the target opacity to high-energy cosmic rays spans several orders of magnitude. 

Figure~\ref{fig:density_structure}(b) illustrates the cumulative column density, $\Sigma(r)$, integrated inward from the stellar surface. Defining the characteristic hadronic interaction length as $\lambda_{pp} \approx 50\ \mathrm{g\ cm^{-2}}$, the results demonstrate that the cumulative column density exceeds this threshold at a shallow normalized depth of $(R_* - r)/R_* \sim 10^{-3}$, where the pair-production optical depth $\tau_\gamma$ is already substantial. This quantitatively confirms the severe geometric restriction of observable gamma-ray emission to a thin atmospheric skin, directly reflecting the exponential attenuation operators formulated in Sec.~\ref{sec:magnetic_mirroring}.

The assumed cosmic-ray propagation mechanism strongly dictates the multi-messenger yields. Figure~\ref{fig:multimessenger_spectra} compares the single-star spectra of the Sun under the field-free ballistic limit against the magnetic mirroring scenario. Under the ballistic approximation, the predicted TeV gamma-ray luminosity significantly underpredicts the HAWC detection. In contrast, introducing magnetic mirroring—driven by horizontal internetwork fields—substantially elongates the effective path length of TeV cosmic rays within the photosphere. This brings the predicted spectrum into agreement with the HAWC data, validating the local magnetic modulation framework and providing an empirical calibration for the macroscopic Galactic yield integration.

Figure~\ref{fig:multimessenger_spectra} compares our neutrino predictions with prior studies~\cite{Edsjo2017, Ng2017}. Our ballistic scenario provides a conservative benchmark for the field-free case, yielding results broadly consistent with early estimates~\cite{Edsjo2017}. In contrast, the magnetic mirroring model predicts a suppressed flux that falls below the semi-analytic results of Ref.~\cite{Ng2017}. This discrepancy highlights the role of sub-photospheric magnetic confinement in reducing the integrated stellar neutrino yield relative to standard approximations. Consequently, objects with vast internal mass reservoirs (e.g., massive stars and red giants) naturally dominate the macroscopic yield, because they lack such efficient magnetic mirroring, allowing incident cosmic rays to deeply penetrate and utilize their entire massive interiors as transparent targets for neutrino production.

Expanding from this solar benchmark, we systematically compute the emission properties across the full stellar parameter space. Figure~\ref{fig:lum_mass_relation} presents the time-weighted average gamma-ray luminosity ($1\text{--}100\;\mathrm{TeV}$) as a function of initial mass and metallicity. The results reveal two evolutionary regimes governed by fundamentally distinct physical mechanisms.

In the magnetic modulation regime ($M \le 1.41\,M_\odot$), the mean stellar luminosity is anti-correlated with mass (Fig.~\ref{fig:lum_mass_relation}a). For low-mass main-sequence stars, the local dynamo mechanism—sustained by fully or deeply convective envelopes—generates strong surface magnetic fields. This mirroring effect dramatically amplifies the cosmic-ray interaction probability, allowing low-mass stars to achieve mean luminosities exceeding those of solar-type stars and partially compensating for their extremely small geometric areas. In this regime, metallicity exerts a negligible effect on the final yield.

Conversely, massive and late-stage populations occupy a geometric-area-dominated regime (Fig.~\ref{fig:lum_mass_relation}b). As stars evolve off the main sequence, envelope expansion dampens the surface dynamo, reverting particle transport to the ballistic limit. Here, the time-weighted average luminosity is driven entirely by the dramatic radial expansion during late evolutionary stages. Notably, the high-mass yield ($M \gtrsim 15\,M_\odot$) is highly sensitive to metallicity: high-metallicity models ($Z \ge 0.006$) experience intense line-driven winds and severe envelope inflation, maintaining asymptotic luminosities $> 10^{20}\ \mathrm{TeV\ s^{-1}}$. Low-metallicity stars, by contrast, retain relatively compact envelopes, strictly limiting their geometric target area and suppressing their total emission.

\subsection{Spatial Distribution Morphology of the Galactic Stellar Diffuse Background}

\begin{figure*}[htbp]
    \centering
    \begin{minipage}{0.8\textwidth}
        \centering
        \includegraphics[width=\linewidth]{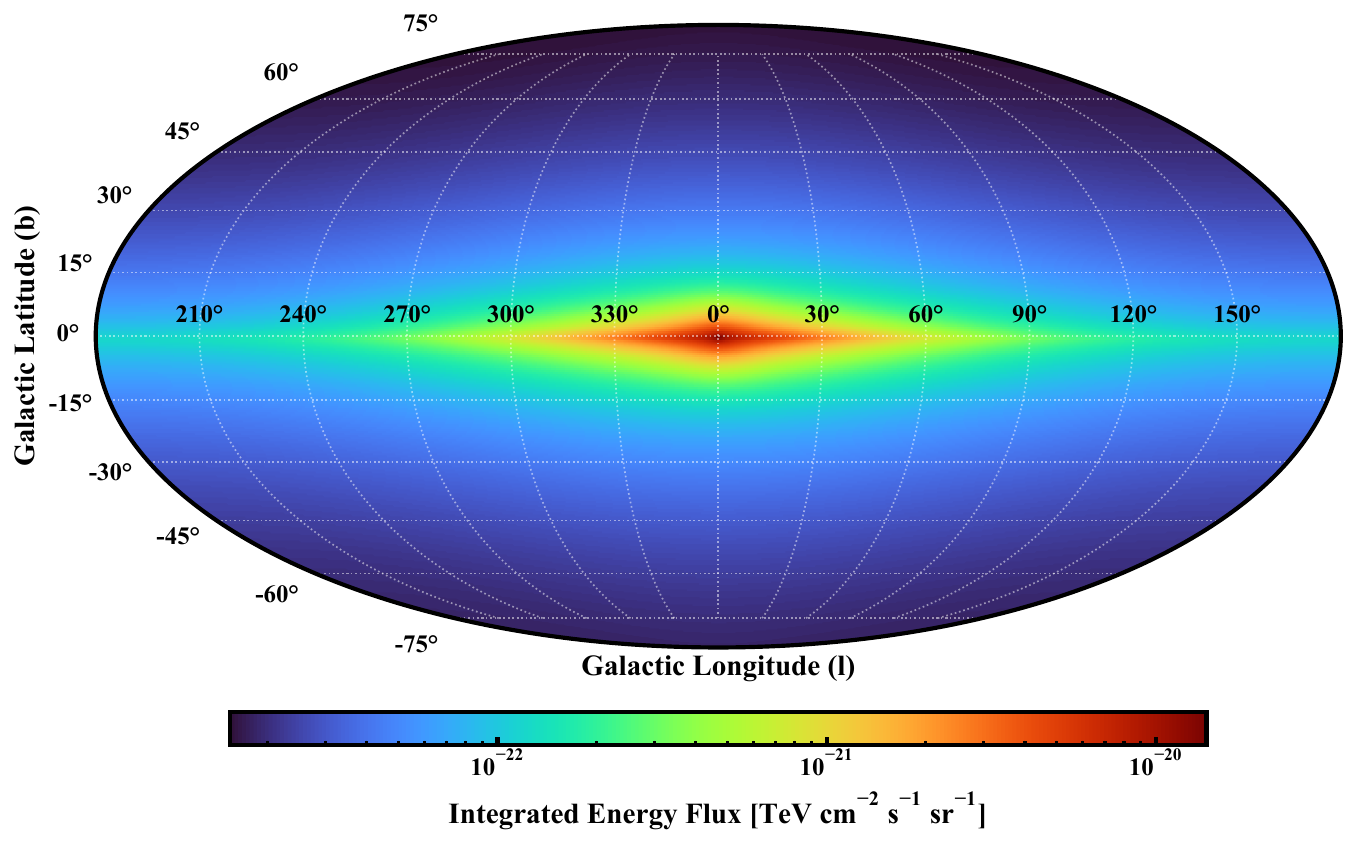}
        \textbf{(a) Gamma-ray Energy Flux}
    \end{minipage}
    \vspace{0.6cm}
    \begin{minipage}{0.8\textwidth}
        \centering
        \includegraphics[width=\linewidth]{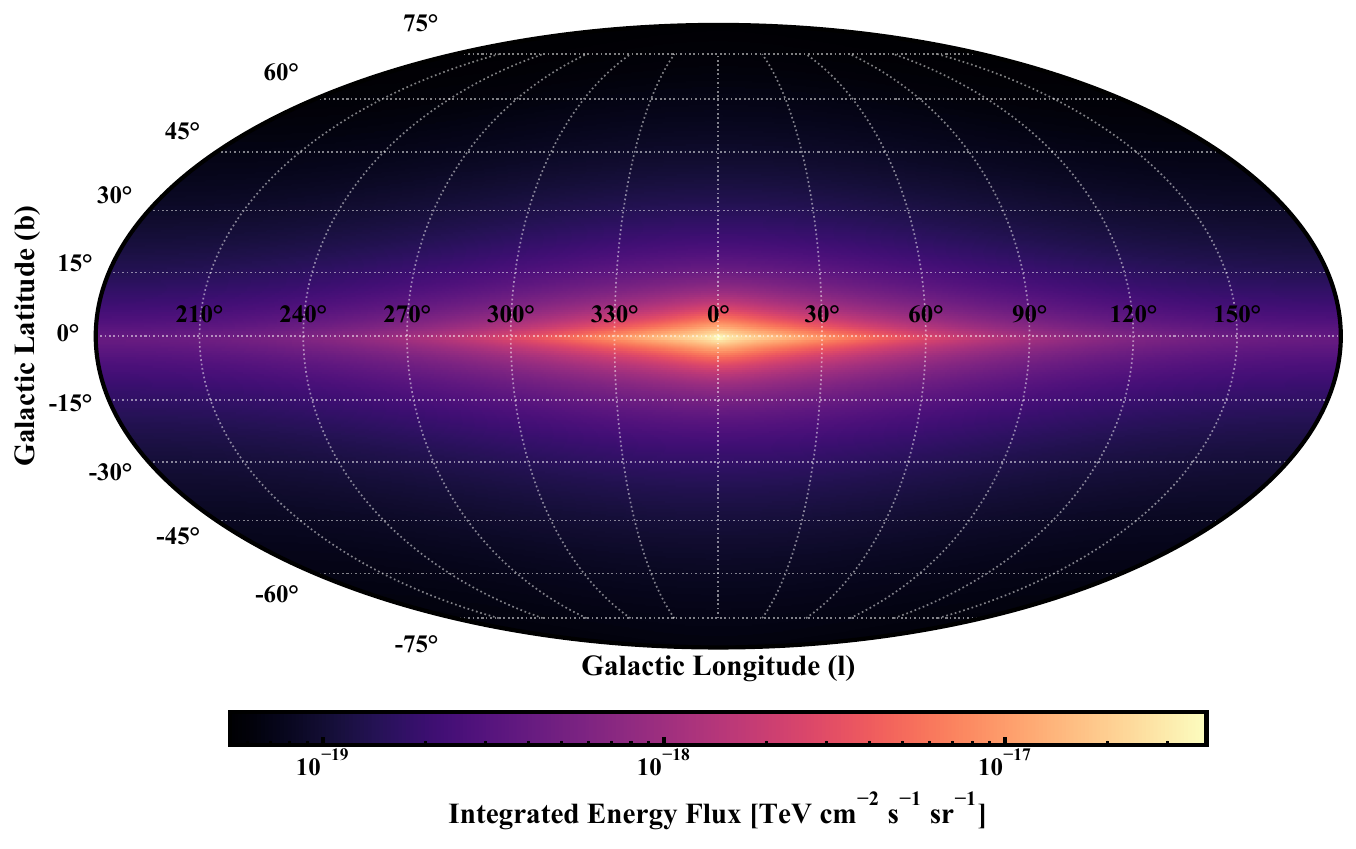}
        \textbf{(b) Neutrino Energy Flux}
    \end{minipage}
    \caption{All-sky intensity maps of the predicted multi-messenger emission from Galactic stellar atmospheres ($E > 1.0\;\mathrm{TeV}$) in Galactic coordinates (Mollweide projection). Panels (a) and (b) display the integrated energy flux for gamma rays and neutrinos, respectively. The emission morphology is tightly coupled to the Galactic stellar disk and bulge components.}
    \label{fig:skymaps}
\end{figure*}
We scale the microscopic single-star yields to macroscopic observables via full-Galaxy population synthesis. Figure~\ref{fig:skymaps} presents the predicted all-sky integrated energy flux distributions for gamma rays and neutrinos at $E > 1.0\ \mathrm{TeV}$. The diffuse multi-messenger emission from stellar atmospheres is strongly concentrated toward the Galactic Center, tracing the underlying stellar number density profiles of the bulge and disks. Compared to conventional diffuse background models based on interstellar gas, the stellar background displays a sharper contrast near the Galactic plane ($b = 0^\circ$), and its high-latitude attenuation closely maps the exponential scale height of the stellar disk. This highly flattened morphology provides a distinct theoretical template for the spatial distribution of stellar multi-messenger emission.

\begin{figure}[t]
    \centering
    \includegraphics[width=0.95\columnwidth]{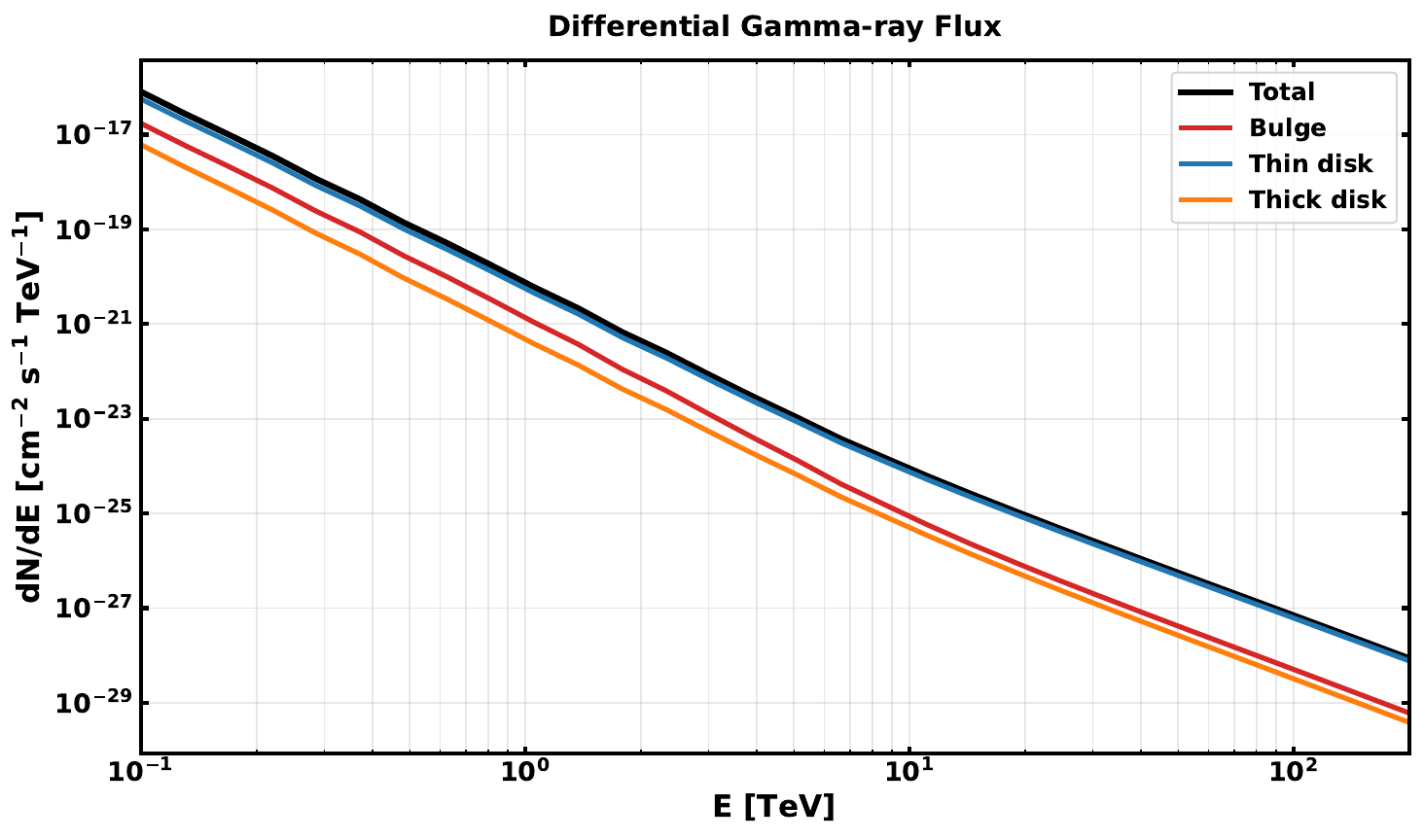}
    \caption{Component-resolved total gamma-ray differential flux \(dN/dE\) expected from the bulge, thin-disk, and thick-disk components, in units of \({\rm cm^{-2}\,s^{-1}\,TeV^{-1}}\).}
    \label{fig:gamma_component_dnde}
\end{figure}

We decompose the total gamma-ray flux into the differential spectra, $dN/dE$, of the bulge, thin disk, and thick disk (Fig.~\ref{fig:gamma_component_dnde}). Across the entire energy range, the thin disk overwhelmingly dominates the emission, with minor contributions from the bulge and the thick disk.

\subsection{Quantitative Comparison with Experimental Observations and Physical Constraints}

\begin{figure*}[htbp]
    \centering
    \begin{minipage}{0.32\textwidth}
        \centering
        \includegraphics[width=\linewidth]{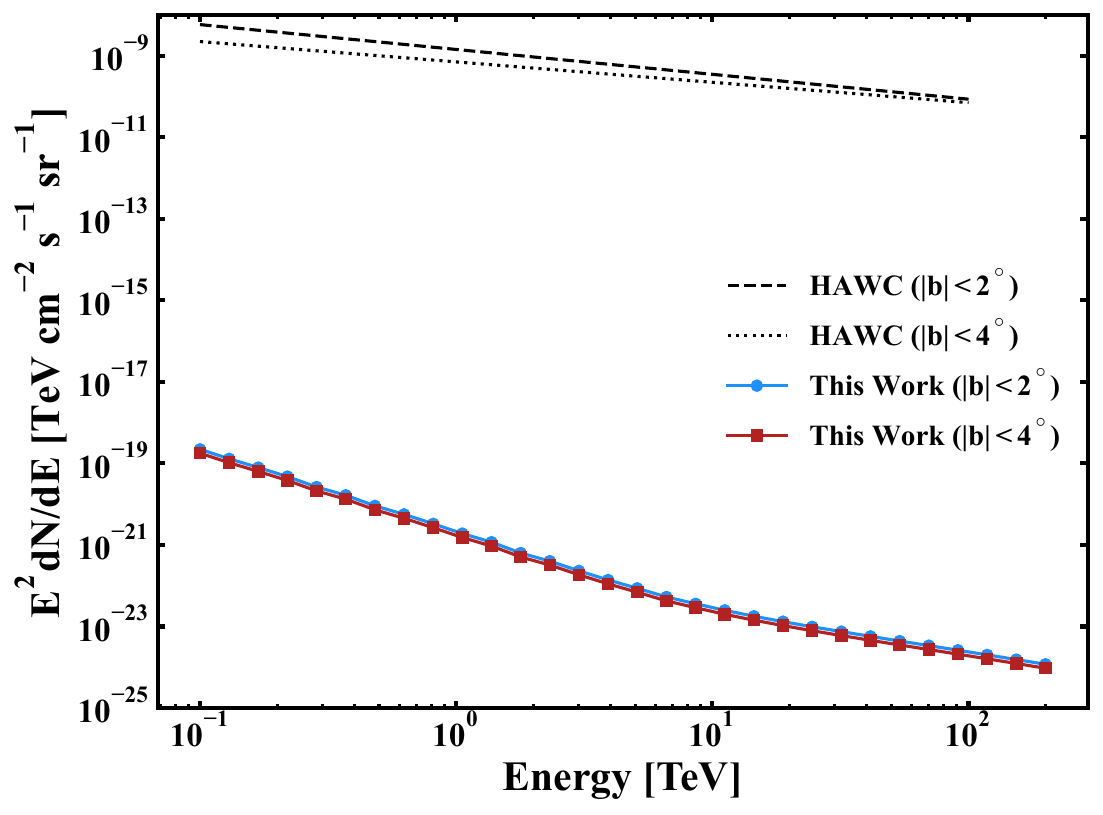}
        \textbf{(a) HAWC Region}
    \end{minipage}
    \hfill
    \begin{minipage}{0.32\textwidth}
        \centering
        \includegraphics[width=\linewidth]{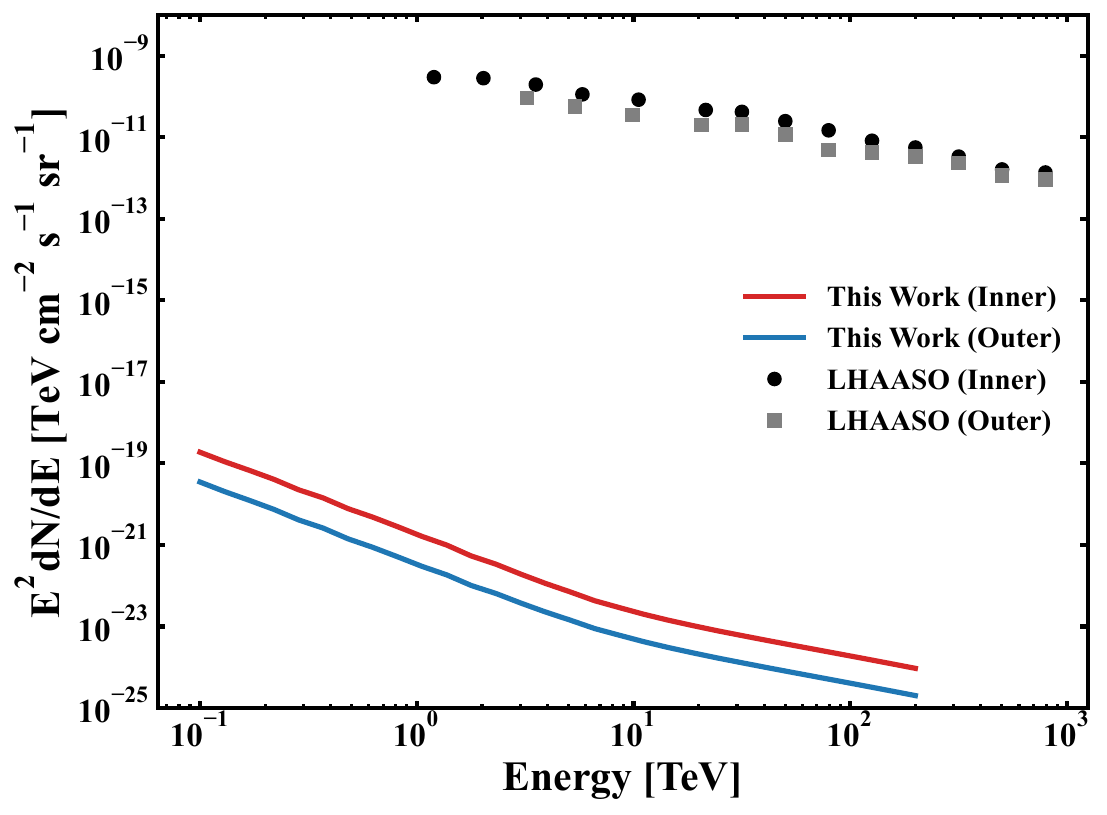}
        \textbf{(b) LHAASO Region}
    \end{minipage}
    \hfill
    \begin{minipage}{0.32\textwidth}
        \centering
        \includegraphics[width=\linewidth]{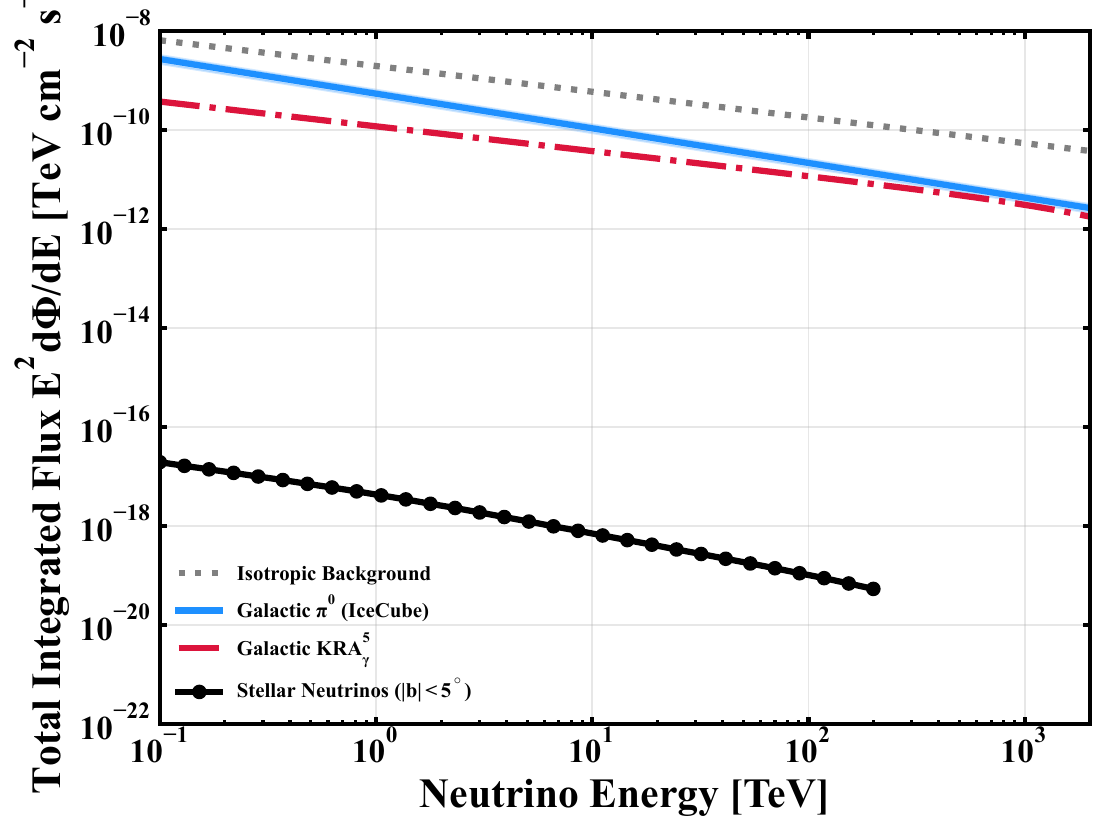}
        \textbf{(c) Galactic Plane Neutrino}
    \end{minipage}
    \caption{
    Comparison of the predicted stellar diffuse background with experimental data in the Galactic plane.
    (a) and (b) show the gamma-ray flux relative to HAWC and LHAASO measurements across different Galactic latitude intervals.
    (c) Integrated stellar neutrino flux ($|b| < 5^\circ$, black solid) compared against IceCube observations and the standard KRA$_\gamma$ model.
        Across the probed energy range, the predicted stellar contributions are vanishingly small. The stellar gamma-ray and neutrino fluxes fall up to $\sim 10$ and $\sim 7$--$8$ orders of magnitude below the measured diffuse emissions, respectively, confirming they are negligible sources of the Galactic high-energy diffuse flux.}
    
    \label{fig:comparison_low_lat}
\end{figure*}

\begin{figure*}[htbp]
    \centering
    \begin{minipage}{0.48\textwidth}
        \centering
        \includegraphics[width=\linewidth]{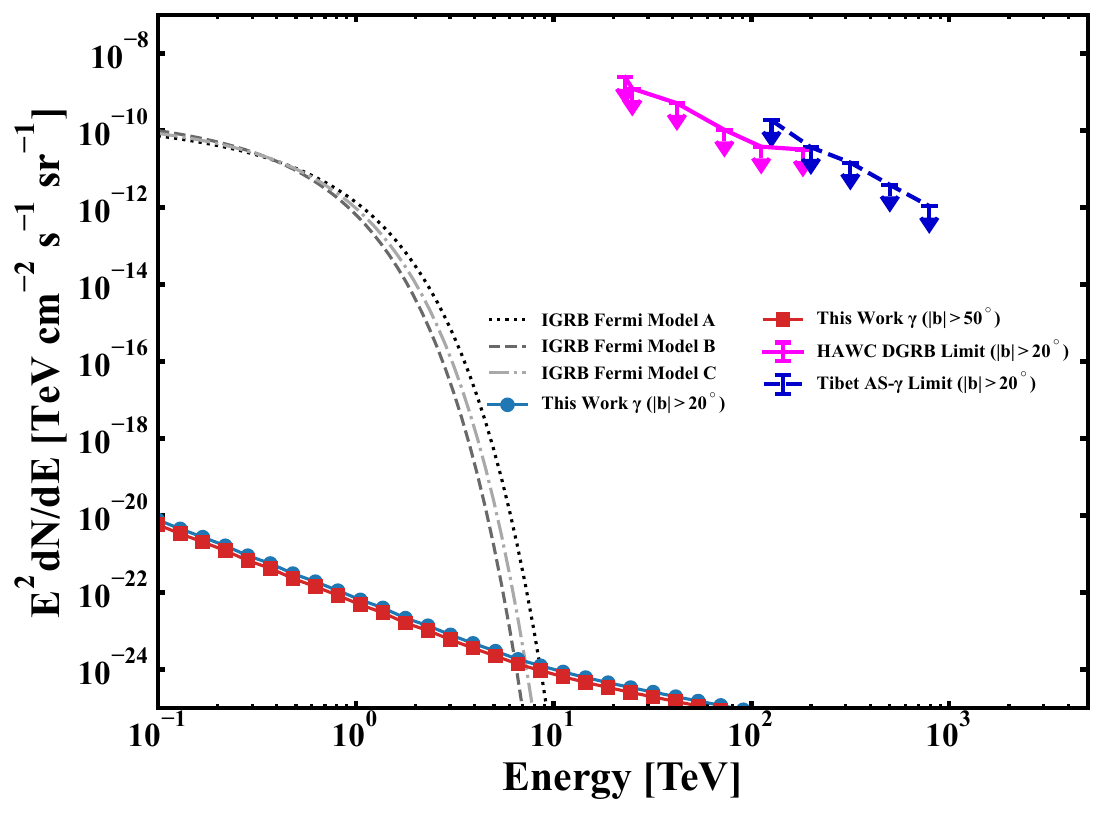}
        \textbf{(a) High-Latitude Gamma-Ray (IGRB)}
    \end{minipage}
    \hfill
    \begin{minipage}{0.48\textwidth}
        \centering
        \includegraphics[width=\linewidth]{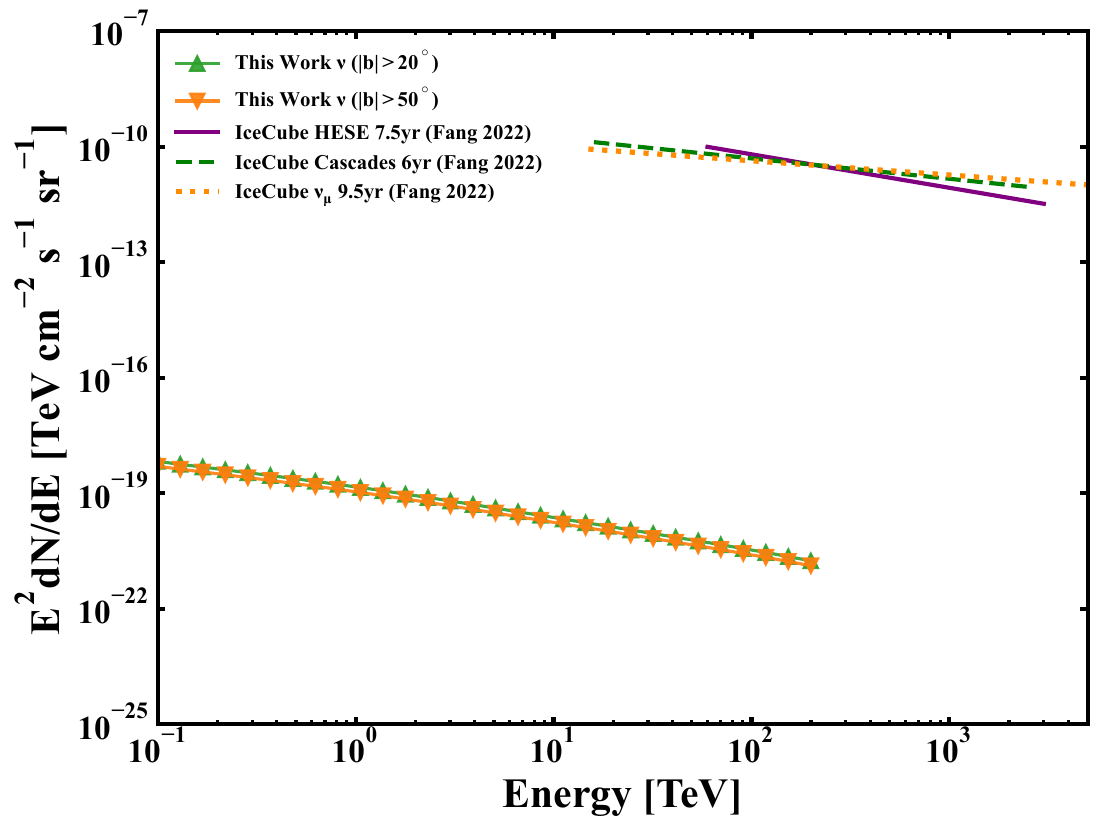}
        \textbf{(b) High-Latitude Neutrino}
    \end{minipage}
    \caption{High-latitude ($|b| > 20^\circ$ and $|b| > 50^\circ$) diffuse emission components compared with experimental limits. (a) Stellar gamma-ray flux vs. the Fermi-LAT Isotropic Gamma-Ray Background (IGRB) and VHE upper limits. (b) Stellar neutrino flux vs. IceCube HESE and track-like measurements. The stellar signal provides a hard, irreducible local background that overtakes the EBL-attenuated IGRB at $E > 10\;\mathrm{TeV}$.}
    \label{fig:comparison_high_lat}
\end{figure*}

We quantitatively compare the synthesized stellar atmospheric contribution against high-energy observational data across three key domains. 

First, regarding Galactic plane gamma-ray constraints, Figure~\ref{fig:comparison_low_lat} compares our predicted spectra against the standard observation windows of HAWC~\cite{HAWC_Diffuse2024} and LHAASO~\cite{LHAASO2023}. In the 0.1–1000 TeV band, the collective stellar gamma-ray flux is approximately ten orders of magnitude below the total diffuse emission observed by these arrays. Even toward the Galactic Center, where the stellar density peaks, the contribution remains well below current detector sensitivities. 

Second, for the neutrino channel, Figure~\ref{fig:comparison_low_lat}(c) compares the integrated stellar neutrino flux in the Galactic plane ($|b| < 5^\circ$) against IceCube observations and the standard KRA$_\gamma$ model~\cite{IceCube2023}. The predicted stellar neutrino flux at 1 TeV is approximately $10^{-17}\ \mathrm{TeV\ cm^{-2}\ s^{-1}}$, remaining orders of magnitude below the dominant hadronic ISM backgrounds.

Finally, we address the high-latitude diffuse emission in Figure~\ref{fig:comparison_high_lat}. In the GeV–TeV band, the predicted stellar background is far below the Isotropic Gamma-Ray Background (IGRB) upper limits defined by Fermi-LAT. At ultra-high energies ($E > 10\ \mathrm{TeV}$), the expected flux from collective stellar point sources formally exceeds the extrapolated power-law Fermi IGRB model. However, this crossover is primarily driven by severe Extragalactic Background Light (EBL) attenuation, which exponentially suppresses true extragalactic photons at $>10\ \mathrm{TeV}$, rather than an absolute dominance of the stellar signal. While this stellar contribution introduces a novel, irreducible local physical component to the UHE high-latitude sky, it currently remains below the detection thresholds of ground-based arrays.

\section{Discussion and Conclusion}
\label{sec:discussion_and_conclusion}

In this work, we systematically evaluate the cumulative multi-messenger diffuse emission from the Galactic stellar ensemble by coupling MESA structural profiles and magnetic-field-modulated cosmic-ray transport with a 3D population synthesis model. We demonstrate that the collective gamma-ray emission from stellar atmospheres constitutes a physically negligible background. Excluding the quiescent Sun, the aggregate stellar flux remains orders of magnitude below contemporary instrument sensitivities. This suppression is governed by the minimal geometric filling factor of the stellar population and severe exponential pair-production attenuation ($e^{-\tau_\gamma}$).

We adopt an axisymmetric Galactic model, as spiral arms primarily redistribute stars azimuthally without altering the total disk mass. Incorporating local density contrasts from Gaia DR3 arm loci [34--37] perturbs the predicted gamma-ray and neutrino fluxes by merely $3\%$--$6\%$ at low latitudes and $< 1\%$ at high latitudes. Therefore, non-axisymmetric structures are strictly secondary effects, and our robust conclusion remains unchanged: the stellar atmospheric contribution remains well below current observational sensitivity and does not require an additional background subtraction.


This definitive constraint independently validates the robustness of standard interstellar medium (ISM) diffuse background models. Because the cumulative stellar gamma-ray flux is vastly subdominant to the conventional hadronic ISM background, it does not introduce unmodeled systematic uncertainties. Consequently, future searches for Galactic PeVatrons or dark matter annihilation signatures can proceed without requiring complex stellar subtraction templates.

In the neutrino channel, the entire stellar interior effectively acts as an optically thin target, yet the macroscopic yield remains highly subdominant. The diffuse 1\,TeV neutrino intensity ($\sim 10^{-17}\;\mathrm{TeV\;cm^{-2}\;s^{-1}\;sr^{-1}}$) is $\sim7$--$8$ of magnitude below current IceCube observations. This emission is safely encompassed within the statistical uncertainties of conventional hadronic ISM models (e.g., KRA$_\gamma$) at low latitudes and remains well below the isotropic background at high latitudes.

The sole exception emerges at ultra-high energies ($E > 10$\,TeV). Here, severe extragalactic background light (EBL) attenuation suppresses the extragalactic isotropic gamma-ray background (IGRB), allowing the unattenuated local stellar flux to establish an irreducible physical floor for next-generation observatories. Finally, as our quantitative limits are tied to the phenomenological parameterization of stellar magnetospheres ($f_B$ and $f_{struct}$), refined magnetohydrodynamic models will be pivotal in further constraining this local astrophysical background.

\noindent\textbf{Acknowledgments.}
This research work is also supported by the following grants: the National SKA Program of China (2025SKA0110104), the National Natural Science Foundation of China NSFC (12273114), the Project for Young Scientists in Basic Research of the Chinese Academy of Sciences (YSBR-061), the Jiangsu Innovation and Entrepreneurship Talent Team Program (JSSCTD202436), and the Youth Fund of the Basic Research Program of Jiangsu (Grant No. BK20251706).


\end{document}